\documentclass[%
 reprint,
superscriptaddress,
 amsmath,amssymb,
 aps,
pre,
floatfix
]{revtex4-2}
\usepackage{caption}
\usepackage{graphicx}
\usepackage{dcolumn}
\usepackage{bm}
\usepackage{xcolor}
\usepackage{subcaption}
\usepackage[colorlinks=true,
            linkcolor=black,
            citecolor=blue,
            urlcolor=black]{hyperref}

\allowdisplaybreaks

\begin{document}

\preprint{APS/123-QED}

\title{Beyond binary scission: a generalized three-species cascade breakage model for wormlike micellar solutions}

\author{Rongxin Lu}
\affiliation{School of Mathematics, Jilin University, Changchun 130012, China}
\author{Jiwei Jia}
\email[Corresponding author: ]{jiajiwei@jlu.edu.cn} 
\affiliation{School of Mathematics, Jilin University, Changchun 130012, China}
\affiliation{AI for Science and Engineering Center, Shenzhen Loop Area Institute, Shenzhen 518048, China}
\author{Young Ju Lee}
\email[Corresponding author: ]{yjlee@txstate.edu} 
\affiliation{Department of Mathematics, Texas State University, San Marcos, TX 78666, USA}






\begin{abstract}
Wormlike micellar fluids exhibit complex rheological behavior driven by the continuous breakage and recombination of self-assembled micellar networks. Existing two-species models provide a coarse binary representation of the micellar population, limiting their ability to resolve intermediate structural states and broad relaxation spectra. To address this limitation, we develop a three-species cascade breakage model consisting of gel-network, long chains, and short chains. By introducing an intermediate micellar state, the model links the rapid relaxation of short fragments to the slow recovery of the gel-network within a unified kinetic framework. This additional structural pathway gives rise to a three-mode viscoelastic response, improves the high-frequency description of the dynamic moduli, and produces a non-monotone constitutive curve that evolves into a stress plateau with coexisting shear bands in Couette flow. This cascade mechanism also governs the transient response, including stress overshoot, hysteresis, and multistep relaxation after shear cessation. Overall, the proposed three-species model provides a physically interpretable framework for wormlike micellar shear banding, capturing the connection between cascade microstructural evolution, broad relaxation dynamics, and macroscopic flow localization.
\end{abstract}

\maketitle


\section{Introduction}

Wormlike micelles are self-assembled, highly flexible cylindrical aggregates formed by amphiphilic surfactant molecules in aqueous solutions \cite{cates1987reptation}. Due to their unique ability to continuously break and reform, wormlike micellar solutions exhibit complex macroscopic rheological behaviors, including viscoelasticity, shear thinning, and shear-induced structural transitions \cite{dreiss2007wormlike, rehage1991viscoelastic,berret1993linear}. These properties make them highly valuable in numerous industrial applications, ranging from enhanced oil recovery to personal care products and drag reduction \cite{yang2002viscoelastic}. One of the most striking rheological signatures of wormlike micellar solutions is their tendency to undergo a shear banding transition \cite{Lerouge2010, olmsted2008perspectives,berret1997inhomogeneous,britton1997two,becu2004spatiotemporal,salmon2003velocity,lettinga2009competition,fardin2016shear}. Within a critical range of shear rates, the initially uniform flow becomes unstable and splits into distinct macroscopic bands. These bands possess different local shear rates and microstructures, allowing the fluid to maintain a nearly constant macroscopic stress.

Theoretical efforts to capture shear banding dynamics have progressively evolved from early phenomenological theories to microstructurally informed network models. Early phenomenological constitutive models, such as the Johnson-Segalman (JS) \cite{johnson1977model} and partially extended convected (PEC) \cite{larson1999} models, successfully produce non-monotonic flow curves but remain mathematically degenerate and lack a direct connection to the underlying chain scission mechanics. The paradigm shift toward physically grounded models was initiated by Cates' reptation-reaction theory \cite{cates1987reptation}, which established the kinetic rules for micellar breakage and recombination. Building upon this microstructural foundation, subsequent theoretical advancements branched into two important directions. To resolve the spatial degeneracy of shear bands, non-local stress diffusion was explicitly incorporated, notably in the diffusive JS (d-JS) model \cite{olmsted2000johnson, radulescu2003time,fielding2006nonlinear,fardin2012instabilities}. Concurrently, to capture the shear-induced structural degradation, kinetic models such as the Bautista-Manero-Puig (BMP) framework \cite{bautista1999understanding} introduced a scalar parameter to phenomenologically track network breakdown. These spatial and structural insights ultimately culminated in the formulation of the Vasquez-Cook-McKinley (VCM) model \cite{Vasquez2007}. By defining a discrete mass exchange between long entangled chains and shorter segments coupled with non-local diffusion, the VCM model provided a comprehensive two-species framework that successfully captured complex transient and steady state banding flows \cite{cromer2011pressure, Zhou2014,dimitriou2012rheo}. Nevertheless, this entire theoretical lineage shares a fundamental limitation. Whether employing a homogenized scalar fluidity or a discrete binary scission assumption, these preceding frameworks do not explicitly resolve the multiscale cascade breakage and the resulting continuous relaxation regimes inherent to real micellar networks.

To bridge the gap between discrete macroscopic network frameworks and the continuous physics of Cates' original theory, we propose a three-species cascade breakage model. Although a population-dynamics-based three-species description has previously been proposed to incorporate gelation and reproduce shear thickening behavior by coupling species populations to an effective viscosity \cite{kang2016three}, a three-species formulation that directly couples cascade scission kinetics with viscoelastic conformation tensors and constitutive stress has not been developed. The classic two-species assumption has clear computational advantages, but it also acts as a first-order discrete approximation of the micellar length distribution and therefore filters out important intermediate structural states. This limitation is closely related to the broad relaxation spectra of reversibly breaking micellar systems \cite{peterson2023wormlike}. In the present model, we introduce three distinct microstructural states: gel-network, entangled long chains, and disentangled short chains, providing a higher-order discrete approximation of the continuous micellar length distribution. By explicitly embedding stress-induced scission mechanisms across these hierarchical kinetic pathways, our model recovers the missing intermediate relaxation modes. This approach captures the multiscale kinetics driving complex transient flows, while keeping the constitutive equations highly tractable for macroscopic simulations.

The remainder of this paper is organized as follows. In Section~\ref{sec-math}, we present the mathematical formulation of the three-species cascade breakage model, including the hierarchical reaction kinetics, the dimensionless constitutive equations, and the flow-dependent scission rates. In Section~\ref{sec-homo}, we study the homogeneous flow predictions of the model, focusing on linear viscoelasticity, steady state shear response, and nonlinear step-strain relaxation. In Section~\ref{sec-inhomo}, we turn to inhomogeneous Couette flow and examine the formation of steady shear-banded states, the transient start-up dynamics, and the multistep relaxation after shear cessation. Finally, we summarize the main findings and discuss the physical implications of the proposed model.

\section{Mathematical formulation}\label{sec-math}
\begin{figure}[htbp]
    \centering
    \includegraphics[height=0.25\textheight]{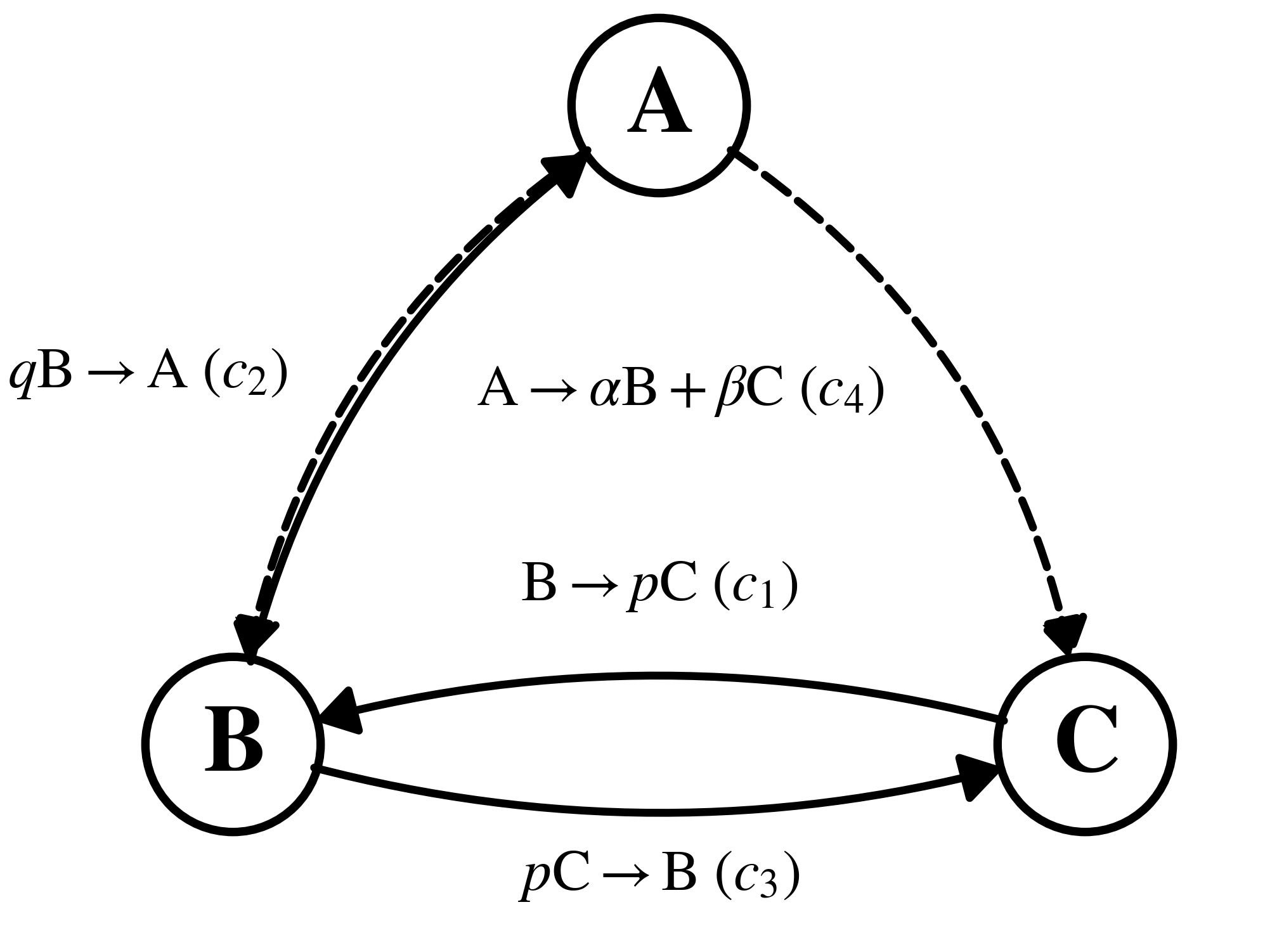}
    \caption{Schematic illustration of the breakage and recombination pathways among the three micellar species: gel-network~A, long chains~B, and short chains~C.}
    \label{fig:species_conversion}
\end{figure}

In this section, we develop a three-species constitutive model to describe the broad relaxation behavior and multiscale breakage processes of wormlike micellar networks under shear \cite{cates1987reptation,peterson2023wormlike}. The micellar network is assumed to consist of three discrete structural states, namely gel-network, long chains, and short chains, which are denoted by species A, B, and C, respectively, as illustrated in Fig.~\ref{fig:species_conversion}.
To rigorously capture the flow-induced microstructural evolution of the micellar network, we adopt a kinetic theory approach based on the continuous phase-space distribution of the constituent chains \cite{bird1987dynamics, beris1994compatibility, bhave1991kinetic, tanaka2010transient}. We define the probability density functions for the three species as $\Psi_\alpha'(\mathbf{r}', \mathbf{Q}', t')$, where $\alpha\in\{A, B, C\}$. The spatiotemporal evolution of these distribution functions is governed by a set of coupled Smoluchowski equations,
\begin{subequations}\label{smol}
\begin{align}
&\frac{\partial \Psi_A'}{\partial t'}
+ \nabla_{\mathbf Q'}\cdot
\left[\left(\mathbf Q'\cdot\nabla\right)\mathbf v_A'\,\Psi_A'\right]
- \nabla_{\mathbf r'}\cdot
\frac{kT}{2\zeta_A}\nabla_{\mathbf r'}\Psi_A'\nonumber\\
&+ \nabla_{\mathbf Q'}\cdot
\frac{2H_A}{\zeta_A}\mathbf Q'\Psi_A'
- \nabla_{\mathbf Q'}\cdot
\frac{2kT}{\zeta_A}\nabla_{\mathbf Q'}\Psi_A'+ \nabla_{\mathbf r'}\cdot\left(\mathbf v_A'\Psi_A'\right)\nonumber \\
&\quad = \frac{1}{q!}c_2'\Psi_B'^q - c_4'\Psi_A',
\\[6pt]
&\frac{\partial \Psi_B'}{\partial t'}
+ \nabla_{\mathbf Q'}\cdot
\left[\left(\mathbf Q'\cdot\nabla\right)\mathbf v_B'\,\Psi_B'\right]
- \nabla_{\mathbf r'}\cdot
\frac{kT}{2\zeta_B}\nabla_{\mathbf r'}\Psi_B'\nonumber\\
&+ \nabla_{\mathbf Q'}\cdot
\frac{2H_B}{\zeta_B}\mathbf Q'\Psi_B'
- \nabla_{\mathbf Q'}\cdot
\frac{2kT}{\zeta_B}\nabla_{\mathbf Q'}\Psi_B'+ \nabla_{\mathbf r'}\cdot\left(\mathbf v_B'\Psi_B'\right) \nonumber\\
&\quad = -c_1'\Psi_B'
-\frac{1}{(q-1)!}c_2'\Psi_B'^q
+\frac{1}{p!}c_3'\Psi_C'^p
+\alpha c_4'\Psi_A',
\\[6pt]
&\frac{\partial \Psi_C'}{\partial t'}
+ \nabla_{\mathbf Q'}\cdot
\left[\left(\mathbf Q'\cdot\nabla\right)\mathbf v_C'\,\Psi_C'\right]
- \nabla_{\mathbf r'}\cdot
\frac{kT}{2\zeta_C}\nabla_{\mathbf r'}\Psi_C' \nonumber\\
&+ \nabla_{\mathbf Q'}\cdot
\frac{2H_C}{\zeta_C}\mathbf Q'\Psi_C'
- \nabla_{\mathbf Q'}\cdot
\frac{2kT}{\zeta_C}\nabla_{\mathbf Q'}\Psi_C' + \nabla_{\mathbf r'}\cdot\left(\mathbf v_C'\Psi_C'\right) \nonumber\\
&\quad = pc_1'\Psi_B'
-\frac{1}{(p-1)!}c_3'\Psi_C'^p
+\beta c_4'\Psi_A'.
\end{align}
\end{subequations}
In the above expressions, $k$ is the Boltzmann constant, $T$ is the absolute temperature, $\zeta_\alpha$ represents the drag coefficient associated with species $\alpha$, and $H_\alpha$ is the entropic spring constant. The source and sink terms on the right-hand side describe the net production and depletion of each species through the cascade scission and recombination processes. To ensure mass conservation across the structural transitions, the strictly positive stoichiometric coefficients $\alpha$ and $\beta$ must satisfy the mass balance relation $p\alpha + \beta = pq$. Following classical transient network theory, we assume a finite spatial extent for the micellar dumbbells \cite{beris1994compatibility, bhave1991kinetic}. This non-local structural assumption gives rise to the migration of the species relative to the bulk flow, leading to the species fluxes $\mathbf{j}_A' = -\frac{kT}{\zeta_A}\nabla'\rho_A' + \frac{pqmH_A}{\zeta_A}\nabla'\cdot\{\mathbf{Q}'\mathbf{Q}'\}_A$, with analogous expressions for $\mathbf{j}_B'$ and $\mathbf{j}_C'$. By integrating Eq.~\eqref{smol} over all configurations $\mathbf{Q}'$, and defining the mass densities as $\rho_A' = 2pqmn_A'$, $\rho_B' = 2pmn_B'$, and $\rho_C' = 2mn_C'$, we recover the material derivatives governing the macroscopic number densities $n_\alpha'$:
\begin{subequations}
\begin{align}
\frac{Dn_A'}{Dt'} &= 2D_A\nabla'^2n_A' - \frac{D_AH_A}{kT}\nabla'\nabla':\{\mathbf{Q}'\mathbf{Q}'\}_A \nonumber\\&\quad+ \frac{1}{q!}c_2'n_B'^q - c_4'n_A', \\[6pt]
\frac{Dn_B'}{Dt'} &= 2D_B\nabla'^2n_B' - \frac{D_BH_B}{kT}\nabla'\nabla':\{\mathbf{Q}'\mathbf{Q}'\}_B \nonumber\\&\quad - c_1'n_B' - \frac{1}{(q-1)!}c_2'n_B'^q + \frac{1}{p!}c_3'n_C'^p + \alpha c_4'n_A', \\[6pt]
\frac{Dn_C'}{Dt'} &= 2D_C\nabla'^2n_C' - \frac{D_CH_C}{kT}\nabla'\nabla':\{\mathbf{Q}'\mathbf{Q}'\}_C \nonumber\\&\quad + pc_1'n_B' - \frac{1}{(p-1)!}c_3'n_C'^p + \beta c_4'n_A',
\end{align}
\end{subequations}
where the translational diffusivities are $D_\alpha = kT / 2\zeta_\alpha$. Similarly, taking the second moment of the distribution by multiplying Eq.~\eqref{smol} by the dyadic product $\mathbf{Q}'\mathbf{Q}'$ and integrating over $\mathbf{Q}'$ leads to the macroscopic evolution equations for the conformation tensors $\{\mathbf{Q}'\mathbf{Q}'\}_\alpha$: 
\begin{subequations}
\begin{align}
    &\{\mathbf{Q}'\mathbf{Q}'\}_{A(1')}+\frac{4H_A}{\zeta_A}\{\mathbf{Q}'\mathbf{Q}'\}_A-\frac{4n_A'kT}{\zeta_A}\mathbf{I}-D_A\nabla'^2\{\mathbf{Q}'\mathbf{Q}'\}_A\nonumber\\
    &=\frac{1}{q!}c_2'\{\mathbf{Q}'\mathbf{Q}'\}_Bn_B'^{q-1}-c_4'\{\mathbf{Q}'\mathbf{Q}'\}_A,\\[6pt]
    &\{\mathbf{Q}'\mathbf{Q}'\}_{B(1')}+\frac{4H_B}{\zeta_B}\{\mathbf{Q}'\mathbf{Q}'\}_B-\frac{4n_B'kT}{\zeta_B}\mathbf{I}-D_B\nabla'^2\{\mathbf{Q}'\mathbf{Q}'\}_B\nonumber\\
    &=-c_1'\{\mathbf{Q}'\mathbf{Q}'\}_B-\frac{1}{(q-1)!}c_2'\{\mathbf{Q}'\mathbf{Q}'\}_Bn_B'^{q-1}\nonumber\\
    &+\frac{1}{p!}c_3'\{\mathbf{Q}'\mathbf{Q}'\}_Cn_C'^{p-1}+\alpha c_4'\{\mathbf{Q}'\mathbf{Q}'\}_A,\\[6pt]
    &\{\mathbf{Q}'\mathbf{Q}'\}_{C(1')}+\frac{4H_C}{\zeta_C}\{\mathbf{Q}'\mathbf{Q}'\}_C-\frac{4n_C'kT}{\zeta_C}\mathbf{I}-D_C\nabla'^2\{\mathbf{Q}'\mathbf{Q}'\}_C\nonumber\\
    &=pc_1'\{\mathbf{Q}'\mathbf{Q}'\}_B-\frac{1}{(p-1)!}c_3'\{\mathbf{Q}'\mathbf{Q}'\}_Cn_C'^{p-1}+\beta c_4'\{\mathbf{Q}'\mathbf{Q}'\}_A,
\end{align}
\end{subequations}
where $(\cdot)_{(1')} = D(\cdot)/Dt' - (\nabla'\mathbf{v}')^T\cdot(\cdot) - (\cdot)\cdot(\nabla'\mathbf{v}')$ denotes the upper convected time derivative, which accounts for the macroscopic rotation and stretching of the continuum fluid element. The total micellar contribution to the stress represents the superposition of the elastic stresses from all three microstructural states: $\bm{\sigma}' = \sum_\alpha H_\alpha\{\mathbf{Q}'\mathbf{Q}'\}_\alpha$.

\subsection{Dimensionless constitutive equations}
For numerical computation and parametric analysis, the governing equations are written in dimensionless form using the following characteristic scales. We scale the spatial coordinate, velocity, and time by $\mathbf{r}=\mathbf{r}'/d, \mathbf{v}=\mathbf{v}'\lambda_{\mathrm{eff}}/d$, and $t=t'/\lambda_{\mathrm{eff}}$, respectively, where $d$ is a characteristic macroscopic length scale and $\lambda_{\mathrm{eff}}$ denotes the effective relaxation time of the self-assembled system. The conformation tensors are normalized by the equilibrium thermal energy and the initial gel density $n_A'^0$, giving
$\mathbf{A}=\frac{H_A\{\mathbf{Q}'\mathbf{Q}'\}_A}{n_A'^0 kT}$, $\mathbf{B}=\frac{H_A\{\mathbf{Q}'\mathbf{Q}'\}_B}{n_A'^0 kT}$, $\mathbf{C}=\frac{H_A\{\mathbf{Q}'\mathbf{Q}'\}_C}{n_A'^0 kT}$, and the normalized number densities $n_\alpha=n_\alpha'/n_A'^0$.

A key feature of the three-species model is the separation of time scales, with each species characterized by a distinct relaxation time $\lambda_\alpha=\zeta_\alpha/4H_\alpha$. We assume that these time scales satisfy $\lambda_A \gg \lambda_B > \lambda_C$, and introduce the relative relaxation ratios $\epsilon_B=\lambda_B/\lambda_A$ and $\epsilon_C=\lambda_C/\lambda_A$, together with the coupling parameters $\mu_A=\lambda_A/\lambda_{\mathrm{eff}}$, $\mu_B=\lambda_B/\lambda_{\mathrm{eff}}$ and $\mu_C=\lambda_C/\lambda_{\mathrm{eff}}$. Following the VCM convention, the dimensionless equilibrium breakage rate of species A is defined as $c_{4,\mathrm{eq}}=\lambda_A c'_{4,\mathrm{eq}}$, where $c'_{4,\mathrm{eq}}$ is the corresponding dimensional rate. Thus $\mu_A=\lambda_A/\lambda_{\mathrm{eff}}=1+c_{4,\mathrm{eq}}$. Assuming that the micellar segments behave as ideal entropic springs \cite{bird1987dynamics,doi1988theory}, the spring constant $H_\alpha$ is inversely proportional to the number of Kuhn segments $N_\alpha$. From the geometric relations $N_A=qN_B=pqN_C$, it follows that the relative elasticities are given by $H_B^*=q$ and $H_C^*=pq$. Therefore, the dimensionless total stress $\bm{\sigma}$, normalized by the plateau modulus $G_0=n_A'^0kT$, takes the form
$\bm{\sigma}=\mathbf{A}+q\mathbf{B}+pq\mathbf{C}$.

Defining the non-dimensional mass diffusivities as $\delta_\alpha = \lambda_A D_\alpha / d^2$, the closed form dimensionless evolution equations for the number densities are:
\begin{subequations}\label{density_dim}
\begin{align}\mu_A\frac{Dn_A}{Dt} &= 2\delta_A\nabla^2n_A - \delta_A\nabla\nabla:\mathbf{A}+ \frac{1}{q!}c_2n_B^q - c_4n_A,\\[6pt]
\mu_A\frac{Dn_B}{Dt} &= 2\delta_B\nabla^2n_B - q\delta_B\nabla\nabla:\mathbf{B} - c_1n_B\nonumber\\&\qquad - \frac{1}{(q-1)!}c_2n_B^q + \frac{1}{p!}c_3n_C^p + \alpha c_4n_A,\\[6pt]
\mu_A\frac{Dn_C}{Dt} &= 2\delta_C\nabla^2n_C - pq\delta_C\nabla\nabla:\mathbf{C} + pc_1n_B \nonumber\\&\qquad- \frac{1}{(p-1)!}c_3n_C^p + \beta c_4n_A,
\end{align}
\end{subequations}
coupled with the dimensionless conformation tensor equations:
\begin{subequations}\label{stress_dim}
\begin{align}
&\mu_A\mathbf{A}_{(1)} + \mathbf{A} - n_A\mathbf{I} - \delta_A\nabla^2\mathbf{A}
= \frac{1}{(q-1)!}c_2n_B^{q-1}\mathbf{B} - c_4\mathbf{A},\\
&\epsilon_B\mu_A\mathbf{B}_{(1)} + \mathbf{B} - \frac{n_B}{q}\mathbf{I}
- \epsilon_B\delta_B\nabla^2\mathbf{B}
=-q\epsilon_Bc_1\mathbf{B}\nonumber \\&\ - \frac{q}{(q-1)!}\epsilon_Bc_2n_B^{q-1}\mathbf{B}
+ \frac{q}{(p-1)!}\epsilon_Bc_3n_C^{p-1}\mathbf{C}
+ \alpha \epsilon_Bc_4 \mathbf{A},\\
&\epsilon_C\mu_A\mathbf{C}_{(1)} + \mathbf{C} - \frac{n_C}{pq}\mathbf{I}
- \epsilon_C\delta_C\nabla^2\mathbf{C}
= pq\epsilon_Cc_1\mathbf{B}\nonumber\\&\ - \frac{pq}{(p-1)!}\epsilon_Cc_3n_C^{p-1}\mathbf{C}
+ \beta \epsilon_Cc_4\mathbf{A}.
\end{align}
\end{subequations}
This microstructural formulation is then coupled to the macroscopic Cauchy momentum equation:
\begin{equation*}
E^{-1}\frac{D\mathbf{v}}{Dt} = \nabla\cdot(-p\mathbf{I} + \beta_s\dot{\bm{\gamma}} + \bm{\tau}), \quad \text{with} \quad \nabla\cdot\mathbf{v}=0.
\end{equation*}
Here, $E = De/Re$ is the elasticity number characterizing the ratio of viscoelastic forces to fluid inertia. The extra stress
\begin{equation*}
    \bm{\tau} = (\mathbf{A}+q\mathbf{B}+pq\mathbf{C})-(n_A+n_B+n_C)\mathbf{I}
\end{equation*}
is supplemented by a Newtonian solvent contribution $\beta_s\dot{\bm{\gamma}}$, where $\beta_s = \eta_s/\eta_0'$, which acts to regularize the flow at extremely high deformation rates.

Crucially, to capture the deformation induced structural degradation of wormlike micelles, we directly couple the scission rates to the local macroscopic flow \cite{larson1984constitutive,marrucci2001integral,likhtman2003simple,turner1992linear, pasquino2025open}. The reformation pathways are assumed to remain close to equilibrium, so that $c_2=c_{2,\mathrm{eq}}$ and $c_3=c_{3,\mathrm{eq}}$. By contrast, the yielding of the gel-network and the subsequent breakage of the long micelles are enhanced by the rate of work imposed by the flow on the elastic segments:
\begin{subequations}
\begin{align}c_1 &= c_{1, \text{eq}} + \frac{1}{3}\xi_1\mu_B\left(\dot{\bm{\gamma}}:\frac{\mathbf{B}}{n_B}\right),\\c_4 &= c_{4, \text{eq}} +\frac{1}{3}\xi_4\mu_A\left(\dot{\bm{\gamma}}:\frac{\mathbf{A}}{n_A}\right).
\end{align}
\end{subequations}
The nonlinear, non-affine retraction parameters $\xi_1$ and $\xi_4$ measure the sensitivity of the corresponding species to the applied velocity gradients. This stress dependent breakage mechanism is responsible for the strong shear thinning response and the subsequent development of distinct shear bands.

To complete the constitutive formulation, we specify the initial conditions from the quiescent equilibrium state. In the absence of macroscopic flow, that is, for $\mathbf{v}=\mathbf{0}$ and $\dot{\bm{\gamma}}=\mathbf{0}$, the stress-induced scission terms vanish, and Eqs.~\eqref{density_dim}–\eqref{stress_dim} reduce to the corresponding equilibrium relations. Since the densities are normalized by the initial gel concentration, we set $n_A^0=1$. The equilibrium number densities of the long and short chains are then determined by the base kinetic rates as follows:
\begin{subequations}
\begin{align*}
n_B^0
&=
\left(\frac{q!\,c_{4,\mathrm{eq}}}{c_{2,\mathrm{eq}}}\right)^{1/q},\\
n_C^0
&=
\left(
\frac{p!\left(qc_{4,\mathrm{eq}}
+c_{1,\mathrm{eq}}n_B^0
-\alpha c_{4,\mathrm{eq}}\right)}
{c_{3,\mathrm{eq}}}
\right)^{1/p},
\end{align*}
\end{subequations}
where $c_{i,\mathrm{eq}}$ represents the dimensionless equilibrium values of the respective reaction rates. Correspondingly, the microstructural conformation tensors relax to their isotropic thermodynamic states, weighted by their equilibrium density fractions:
\begin{equation*}
\mathbf{A}_{\text{eq}}=\mathbf{I}, \quad \mathbf{B}_{\text{eq}}=\frac{n_B^0}{q}\mathbf{I}, \quad \mathbf{C}_{\text{eq}}=\frac{n_C^0}{pq}\mathbf{I}.
\end{equation*}

\section{Homogeneous flow predictions}\label{sec-homo}

To examine the kinetic coupling between structural breakdown and macroscopic stress, we first consider the model under homogeneous flow conditions. Neglecting spatial gradients and nonlocal diffusion, Eqs.~\eqref{density_dim}--\eqref{stress_dim} reduce to a closed system of ordinary differential equations. This simplified setting allows us to study the basic constitutive response of the model before turning to the full spatiotemporal problem of shear banding \cite{Vasquez2007}. In the following computations, we use $p=2$, $q=3$, and $\alpha=\beta=2$.

\subsection{Linear viscoelasticity}
We begin this analysis by considering a small amplitude oscillatory shear (SAOS) perturbation about the quiescent equilibrium state \cite{miller2007transient,pasquino2025open,turner1992linear}. In this regime, the macroscopic response is governed by the effective relaxation time of the network, $\lambda_{\mathrm{eff}}=(\lambda_A^{-1}+c'_{4,\mathrm{eq}})^{-1}$, or equivalently $\lambda_{\mathrm{eff}}=\lambda_A/(1+c_{4,\mathrm{eq}})$ in terms of the dimensionless equilibrium breakage rate. To remain in the linear regime, the imposed harmonic strain, $\gamma_{r\theta}=\mathrm{Re}\{\gamma^0 e^{i\omega t}\}$, is taken to have small amplitude, with $\gamma^0\ll 1$. Here, $\omega=\lambda_{\mathrm{eff}}\omega'$ denotes the dimensionless oscillation frequency. Under this assumption, the flow-induced variations of the scission rates can be neglected, and the reaction rates remain at their equilibrium values. Substituting the harmonic perturbation into the governing equations and retaining only the leading order terms in $\gamma^0$, we obtain the following linearized system for the conformation tensors:
\begin{subequations}\label{saos_linear}
\begin{align}
&i\mu_A\omega A_{r\theta}^1
+ \left(1+c_{4,\mathrm{eq}}\right)A_{r\theta}^1
- \frac{1}{(q-1)!}c_{2,\mathrm{eq}}
(n_B^0)^{q-1}B_{r\theta}^1
\nonumber\\
&\qquad= i\mu_A\omega\gamma^0 A_{rr}^0,
\\[6pt]
&i\epsilon_B\mu_A\omega B_{r\theta}^1
+ \left(
1+q\epsilon_B c_{1,\mathrm{eq}}
+ \frac{q}{(q-1)!}\epsilon_B c_{2,\mathrm{eq}}
(n_B^0)^{q-1}
\right)B_{r\theta}^1
\nonumber\\
&\qquad- \frac{q}{(p-1)!}\epsilon_B c_{3,\mathrm{eq}}
(n_C^0)^{p-1}C_{r\theta}^1
- \alpha\epsilon_B c_{4,\mathrm{eq}}A_{r\theta}^1
\nonumber\\
&\qquad= i\epsilon_B\mu_A\omega\gamma^0 B_{rr}^0,\\[6pt]
&i\epsilon_C\mu_A\omega C_{r\theta}^1
+ \left(
1+\frac{pq}{(p-1)!}\epsilon_C c_{3,\mathrm{eq}}
(n_C^0)^{p-1}
\right)C_{r\theta}^1
\nonumber\\
&\qquad- pq\epsilon_C c_{1,\mathrm{eq}}B_{r\theta}^1
- \beta\epsilon_C c_{4,\mathrm{eq}}A_{r\theta}^1
\nonumber\\
&\qquad= i\epsilon_C\mu_A\omega\gamma^0 C_{rr}^0,\\[6pt]
&(1+c_{4,\mathrm{eq}})A_{rr}^0
- \frac{1}{(q-1)!}c_{2,\mathrm{eq}}(n_B^0)^{q-1}B_{rr}^0
= n_A^0,\\[6pt]
&\biggl(
1+q\epsilon_B c_{1,\mathrm{eq}}
+ \frac{q}{(q-1)!}\epsilon_B c_{2,\mathrm{eq}}
\left(n_B^0\right)^{q-1}
\biggr) B_{rr}^0
= \frac{1}{q}n_B^0\nonumber\\
&\qquad+\frac{q}{(p-1)!}\epsilon_Bc_{3,\mathrm{eq}}
\left(n_C^0\right)^{p-1}C_{rr}^0
+ \alpha\epsilon_B c_{4,\mathrm{eq}}A_{rr}^0,\\[6pt]
&\left(1+\frac{pq}{(p-1)!}\epsilon_C c_{3,\mathrm{eq}}(n_C^0)^{p-1}\right)C_{rr}^0 = pq\epsilon_C c_{1,\mathrm{eq}}B_{rr}^0\nonumber\\
&\qquad+ \beta\epsilon_C c_{4,\mathrm{eq}}A_{rr}^0
+ \frac{1}{pq}n_C^0,\\[6pt]
&\frac{1}{q!}c_{2,\mathrm{eq}}(n_B^0)^q - c_{4,\mathrm{eq}} = 0,\\[6pt]
&-c_{1,\mathrm{eq}}n_B^0 + (\alpha-q)c_{4,\mathrm{eq}}
+ \frac{1}{p!}c_{3,\mathrm{eq}}(n_C^0)^p = 0.
\end{align}
\end{subequations}
In the asymptotic limit associated with the strong separation of time scales, $\lambda_A \gg \lambda_B > \lambda_C$, the coupled system becomes considerably simpler. Retaining terms up to $O(\epsilon)$, the leading order normal components decouple and reduce to their equilibrium values, namely $A_{rr}^0=1$, $B_{rr}^0=n_B^0/q$, and $C_{rr}^0=n_C^0/pq$. As a result, the $O(\gamma^0)$ shear perturbations satisfy a closed linear system for the three-species:
\begin{subequations}
\begin{align}
A_{r\theta}^1=&\gamma^0\biggl(\frac{((\mu_A/(1+c_{4,\mathrm{eq}}))\omega)^2}{1+((\mu_A/(1+c_{4,\mathrm{eq}}))\omega)^2}\nonumber\\
&\qquad+\frac{i(\mu_A/(1+c_{4,\mathrm{eq}}))\omega}{1+((\mu_A/(1+c_{4,\mathrm{eq}}))\omega)^2}\biggr)+o(\gamma^0),\\
B_{r\theta}^1=&\gamma^0\frac{n_B^0}{q}\left(\frac{(\epsilon_B\mu_A\omega)^2}{1+(\epsilon_B\mu_A\omega)^2}+\frac{i\epsilon_B\mu_A\omega}{1+(\epsilon_B\mu_A\omega)^2}\right)+o(\gamma^0),\\
C_{r\theta}^1=&\gamma^0\frac{n_C^0}{pq}\left(\frac{(\epsilon_C\mu_A\omega)^2}{1+(\epsilon_C\mu_A\omega)^2}+\frac{i\epsilon_C\mu_A\omega}{1+(\epsilon_C\mu_A\omega)^2}\right)+o(\gamma^0).
\end{align}
\end{subequations}
By superposing the individual species' contributions to the total stress, the dimensional storage ($G'$) and loss ($G''$) moduli are given by:
\begin{subequations}
\begin{align}
G'(\omega')=&G_0 \biggl\{\frac{(\lambda_{\mathrm{eff}}\omega')^2}{1+(\lambda_{\mathrm{eff}}\omega')^2}+n_B^0\frac{(\lambda_B\omega')^2}{1+(\lambda_B\omega')^2}\nonumber\\
&\qquad\qquad+n_C^0\frac{(\lambda_C\omega')^2}{1+(\lambda_C\omega')^2}\biggr\},\\
G''(\omega')=&G_0\biggl\{\frac{\lambda_{\mathrm{eff}}\omega'}{1+(\lambda_{\mathrm{eff}}\omega')^2}+n_B^0\frac{\lambda_B\omega'}{1+(\lambda_B\omega')^2}\nonumber\\
&\qquad\qquad+n_C^0\frac{\lambda_C\omega'}{1+(\lambda_C\omega')^2}\biggr\}+\eta_s\omega'.
\end{align}
\end{subequations}
\begin{figure}[htbp]
    \centering
    \includegraphics[height=0.25\textheight]{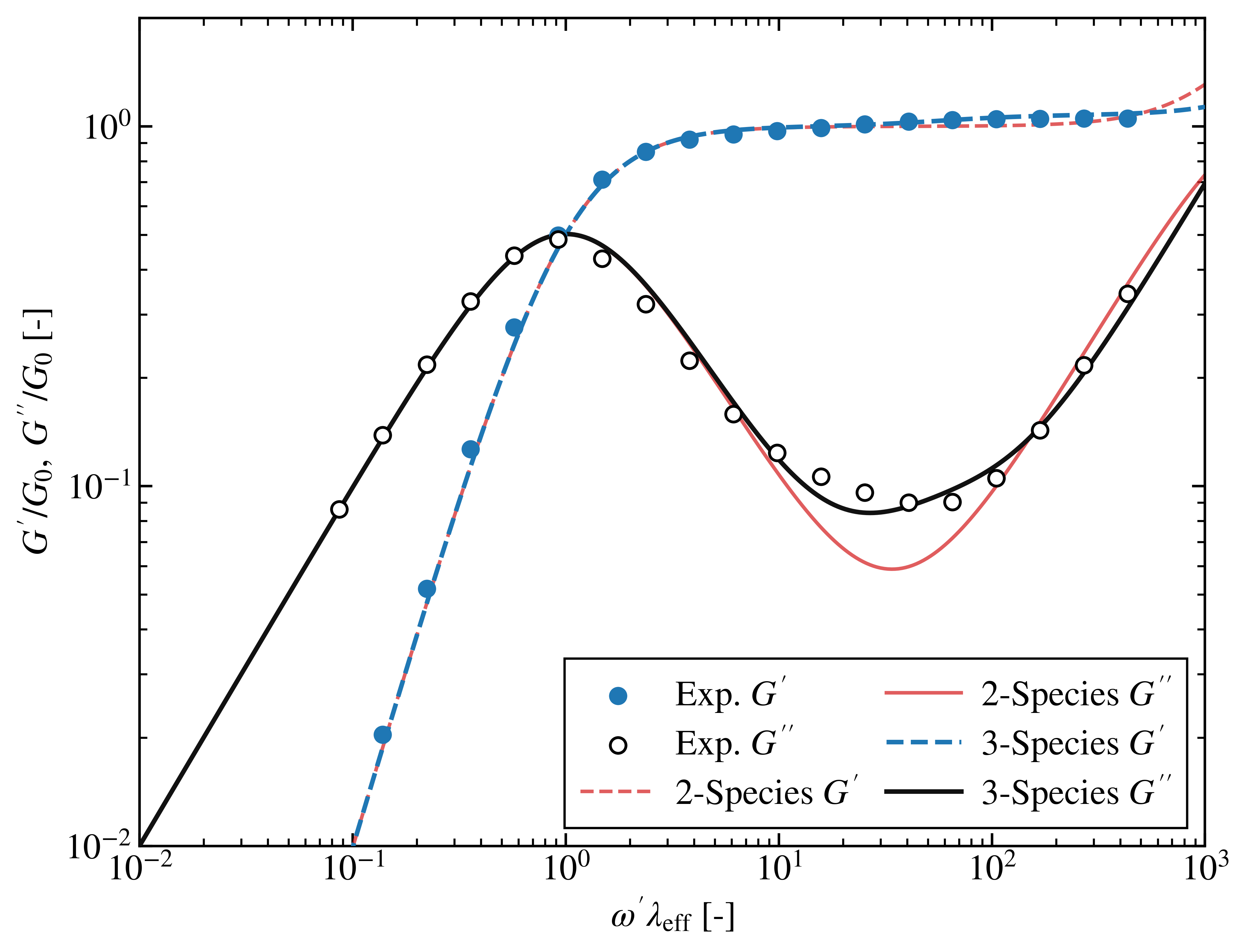}
    \caption{Non-dimensional storage and loss moduli predicted by the two-species and three-species models, compared with experimental master curves for a 100 mM CPyCl/NaSal solution reconstructed from the fitted model parameters~\cite{miller2007transient}.}
    \label{s&l}
\end{figure}

\begin{figure}[htbp]
    \centering

    \begin{subfigure}{\linewidth}
        \centering
        \includegraphics[height=0.25\textheight]{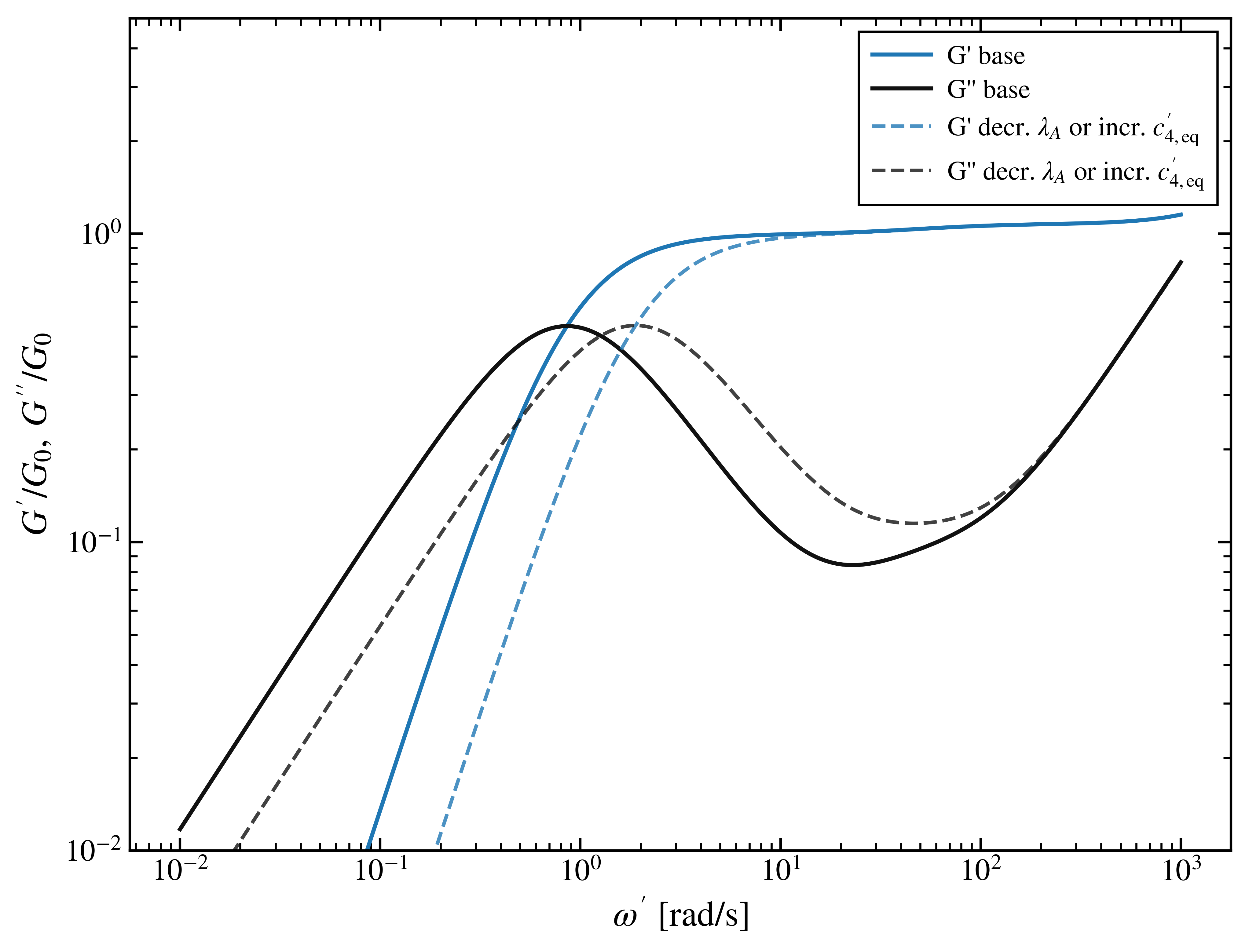}
        \caption{Shifting the primary terminal crossover by decreasing the effective network relaxation time $\lambda_{\mathrm{eff}}$.}
    \end{subfigure}
    \begin{subfigure}{\linewidth}
        \centering
        \includegraphics[height=0.25\textheight]{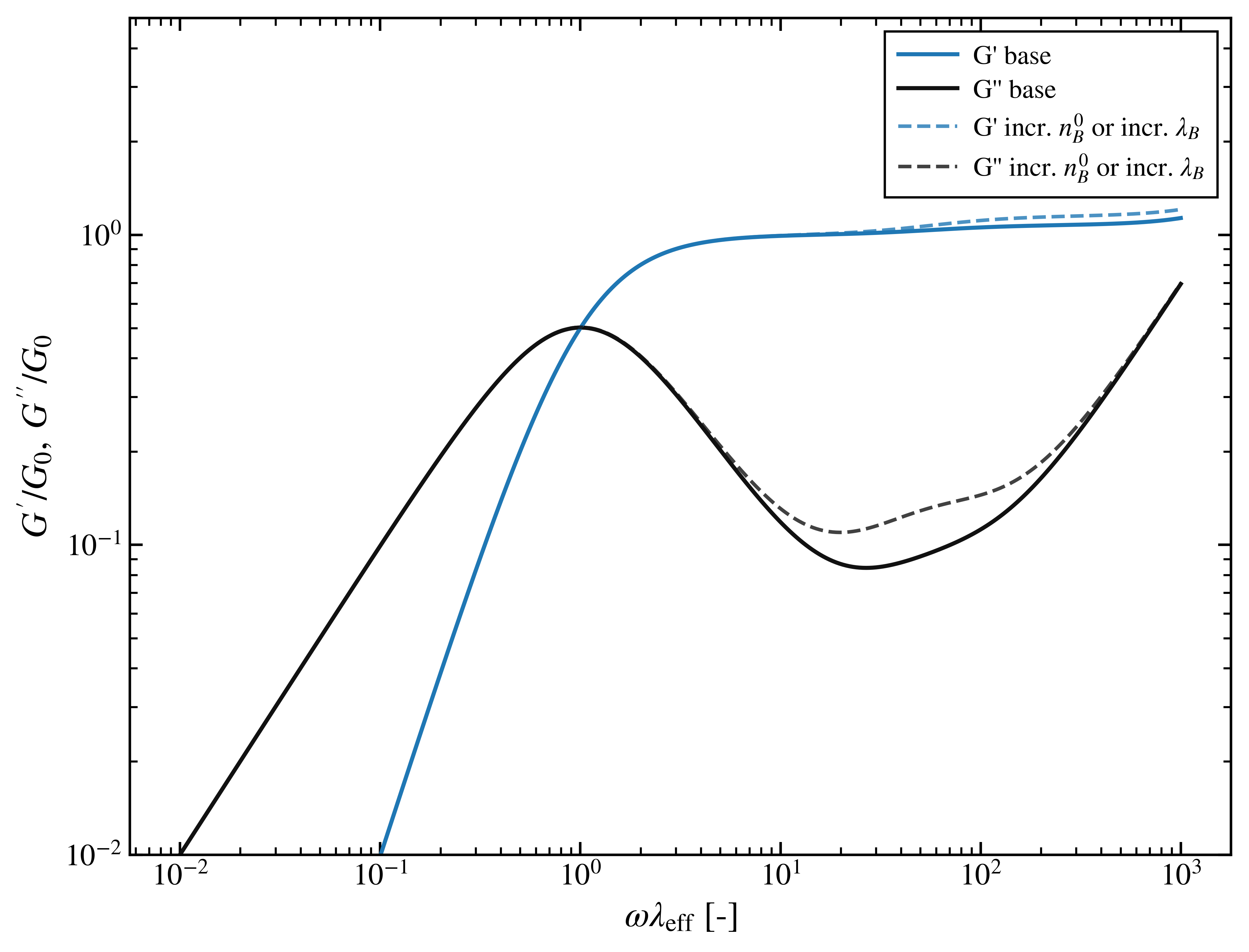}
        \caption{Amplification of the intermediate relaxation regime by increasing the equilibrium density $n_B^0$ or relaxation time $\lambda_B$ of species B.}
    \end{subfigure}
    \begin{subfigure}{\linewidth}
        \centering
        \includegraphics[height=0.25\textheight]{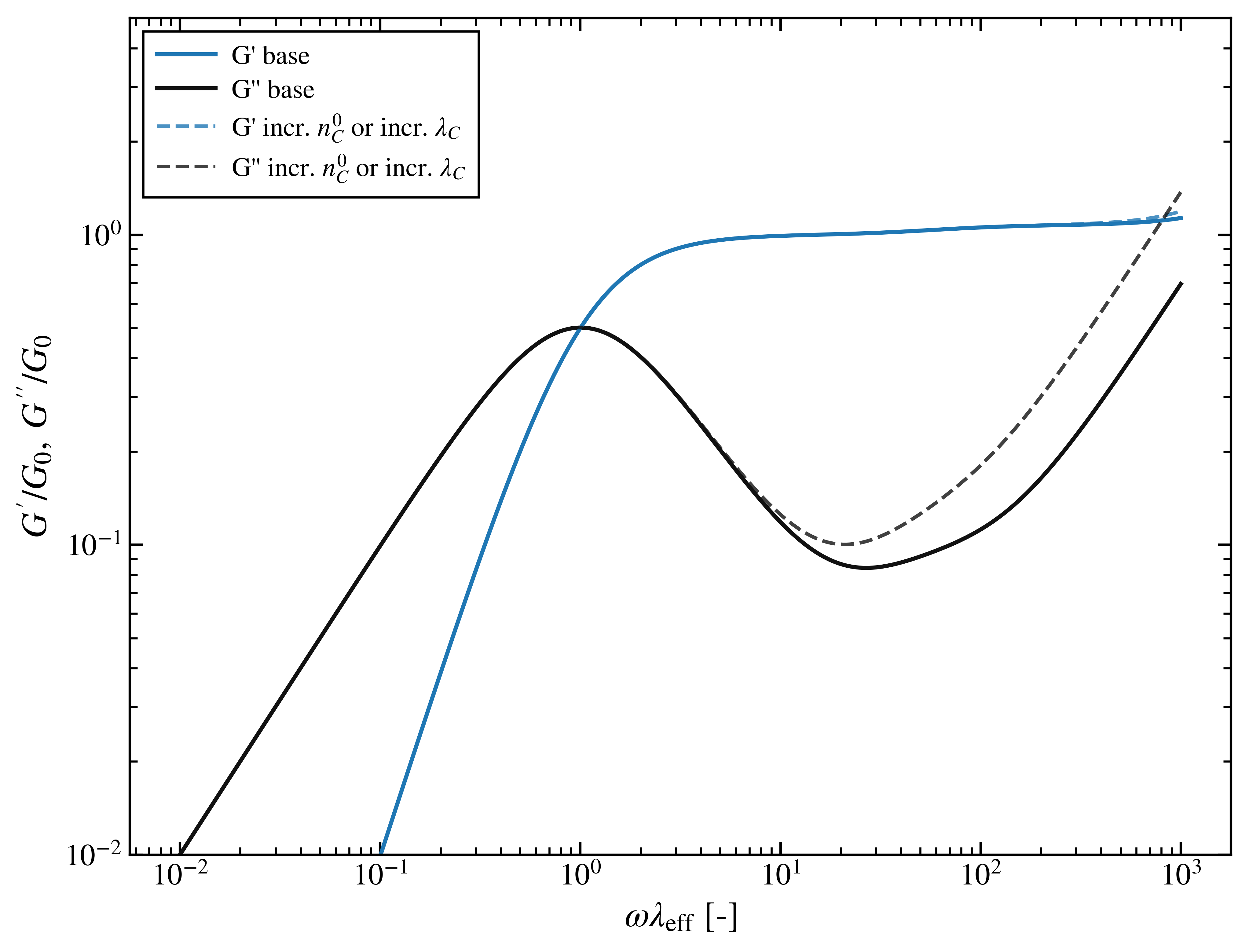}
        \caption{Enhancement of the high frequency Rouse-like response by increasing the density $n_C^0$ or relaxation time $\lambda_C$ of species C.}
    \end{subfigure}
    \caption{Parametric sensitivity of the three-species regimes, isolating the specific kinetic contributions of each micellar component to the dynamic moduli ($G'$ and $G''$).}
    \label{modulus_comparison}
\end{figure}

These equations show that the model admits a discrete three-mode Maxwell structure. In particular, it predicts an upturn in the dynamic moduli at high frequencies, which is consistent with rheological measurements of wormlike micellar solutions.
This behavior mainly comes from the fast viscoelastic relaxation of the intermediate long chains species B and the short chains species C. The final term in the loss modulus corresponds to the Newtonian solvent contribution, which is negligibly small in the present case $(\eta_s/\eta_0'=\beta_s \sim O(10^{-5}))$. Fig.~\ref{s&l} compares the predictions of the two-species and three-species models with the experimental master curves for a 100 mM CPyCl/NaSal solution \cite{miller2007transient}. The experimental data are reconstructed from the fitted model parameters in the literature, with $\lambda_B n_B^0=1.5\times10^{-3}\,\mathrm{s}$, $\lambda_C n_C^0=8\times10^{-4}\,\mathrm{s}$, $\lambda_{\mathrm{eff}}=1.17\,\mathrm{s}$, and $G_0=27\,\mathrm{Pa}$. It can be seen that the three-species model provides a better description of the high frequency regime, where the additional relaxation mode introduced by the third species improves the fit to the upturn in the dynamic moduli. Fig.~\ref{modulus_comparison} further separates the contributions from the three species and illustrates the sensitivity of the three relaxation regimes to the model parameters.
Decreasing the effective relaxation time $\lambda_{\mathrm{eff}}$, either by decreasing the micellar relaxation time $\lambda_A$ or by increasing the baseline breaking rate according to $1/\lambda_{\mathrm{eff}} = 1/\lambda_A + c'_{4,\mathrm{eq}}$, shifts the primary crossover frequency to higher values, as shown in Fig.~\ref{modulus_comparison}a. In addition, increasing the density $n_B^0$ or relaxation time $\lambda_B$ of species B enhances the magnitude of the intermediate second regime, as shown in Fig.~\ref{modulus_comparison}b. By contrast, changes in species C mainly affect the high frequency Rouse-like response in the third regime, shown in Fig.~\ref{modulus_comparison}c.

\subsection{Steady state shear flow}
\begin{figure}[htbp]
    \centering
    \begin{subfigure}{\linewidth}
        \centering
        \includegraphics[height=0.25\textheight]{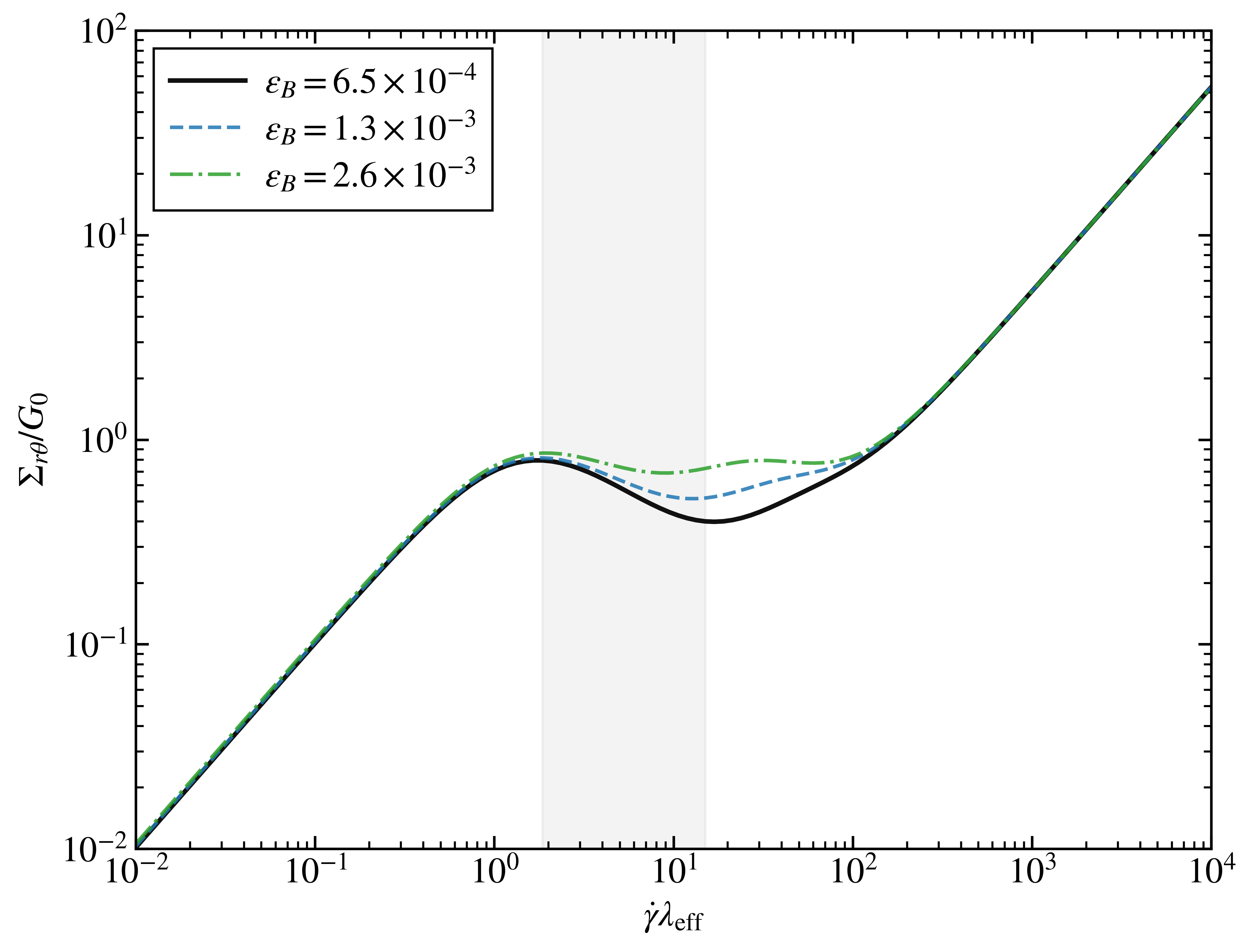}
        \caption{Effect of the dimensionless intermediate relaxation time $\epsilon_B$ on controlling the depth of the unstable viscometric region.}
    \end{subfigure}
    \hfill
    \begin{subfigure}{\linewidth}
        \centering
        \includegraphics[height=0.25\textheight]{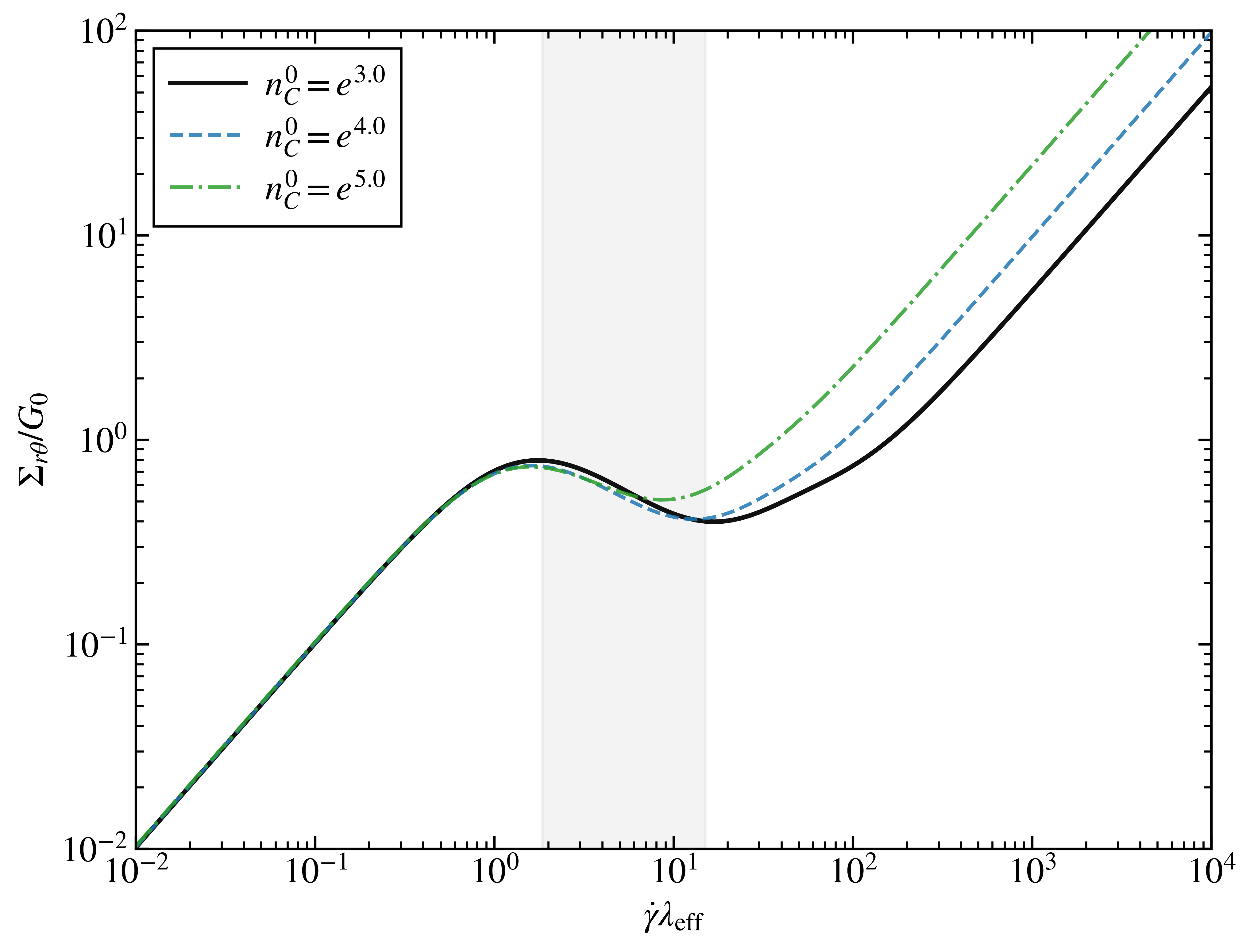}
        \caption{Effect of the short chains density $n_C^0$ on regulating the width of the instability by shifting the high shear viscous upturn.}
    \end{subfigure}

    \caption{Parametric study of the steady state homogeneous flow curve predicted by the three-species model. The shaded area highlights the strictly negative slope, marking the onset of constitutive instability.}
    \label{flow_curve}
\end{figure}

With the linear viscoelastic behavior fully resolved, we proceed to evaluate the nonlinear microstructural evolution under steady homogeneous shear flow. For a steady homogeneous shear flow, $\mathbf{v}=(0,v_\theta(r),0)$ with a spatially uniform shear rate
$\dot{\gamma}=r\partial(v_\theta/r)/\partial r$, explicit time dependence is eliminated, resulting in a fully coupled set of steady state nonlinear equations. In this steady state, the structural integrity of the micellar network is governed by the dynamic balance between shear-induced scission and thermodynamic recombination. Central to this structural transition is the flow enhanced breakage rate $c_4$ of species A. As established previously, this scission rate is fundamentally coupled to the macroscopic deformation via the partially extending strand convection (PEC) mechanism.
During steady shearing, the macroscopic velocity gradient continuously stretches the primary gelation species A. This convective deformation locally elevates the scission rate, thereby forcing a systematic population shift toward the shorter micelles. 
The primary network breaks into intermediate species B, which subsequently undergoes further scission into species C under extreme shear. 
\begin{figure}[htbp]
    \centering
    \begin{subfigure}{\linewidth}
        \centering
        \includegraphics[height=0.25\textheight]{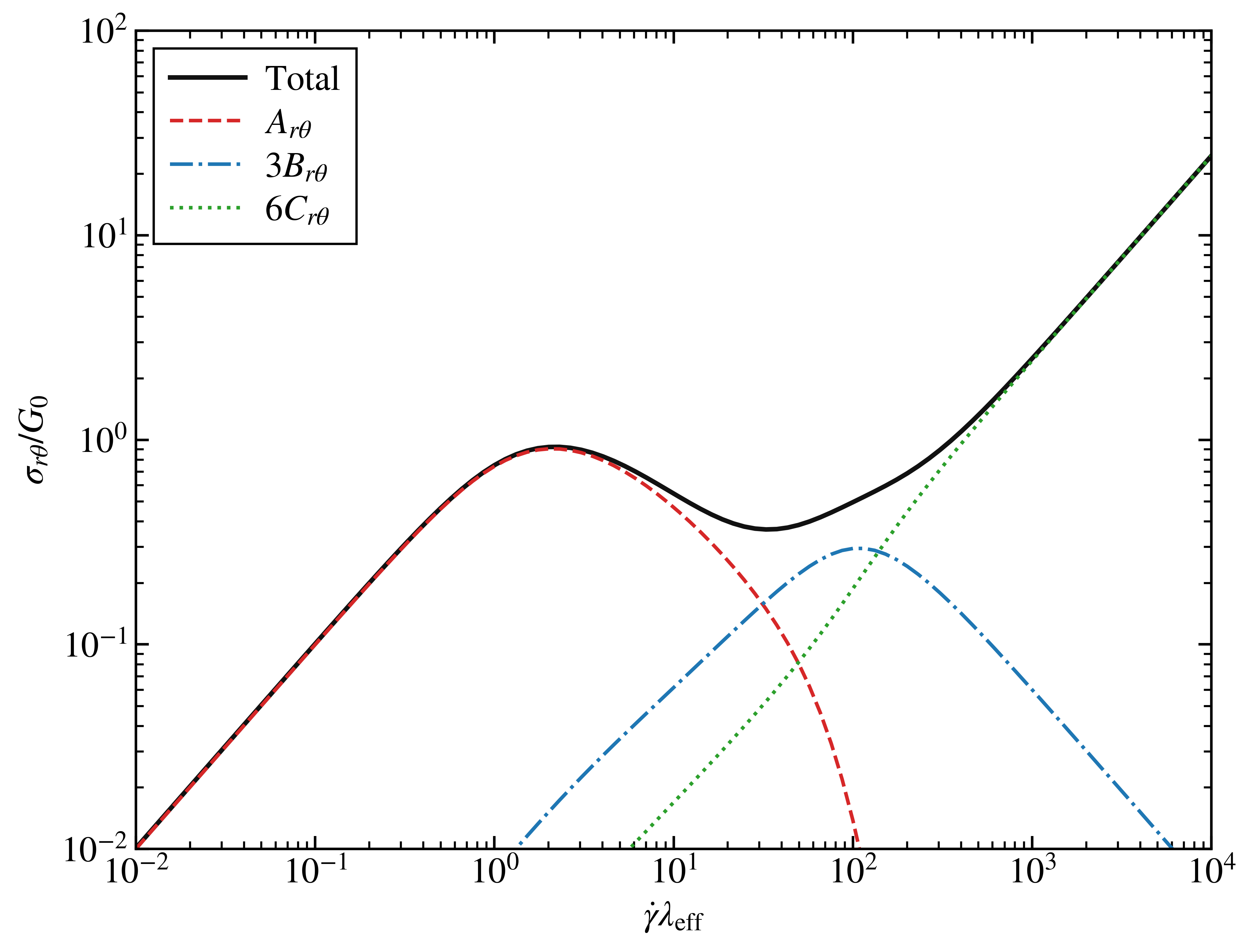}
        \caption{Shear stress.}
    \end{subfigure}
    \begin{subfigure}{\linewidth}
        \centering
        \includegraphics[height=0.25\textheight]{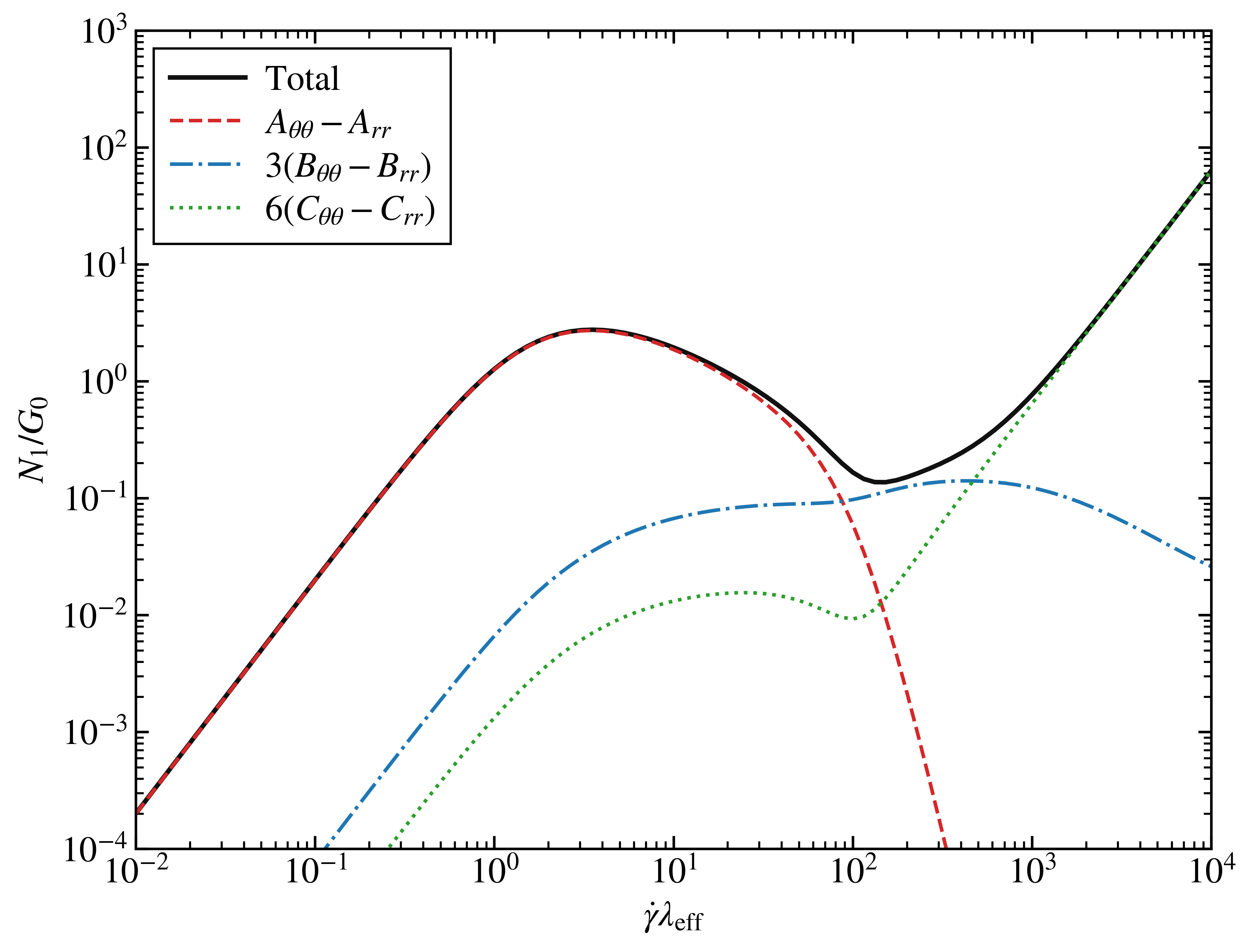}
        \caption{First normal stress difference.}
    \end{subfigure}

    \caption{Model predictions in steady state shear flow as a function of dimensionless shear rate.}
    \label{species_distribution}
\end{figure}
As illustrated in Fig.~\ref{species_distribution}, this continuous redistribution strictly dictates the nonlinear transitions across the three-species regimes.
To map this microstructural evolution onto the macroscopic rheological response, we first evaluate the micellar contribution to the steady shear stress,
\begin{equation*}
    \sigma_{r\theta} = A_{r\theta} + q B_{r\theta} + pq C_{r\theta},
\end{equation*}
and the total viscometric shear stress is then given by
\begin{equation*}
    \Sigma_{r\theta} = \sigma_{r\theta} + \beta_s \dot{\gamma},
\end{equation*}
where the steady state shear components of the conformation tensors $A_{r\theta}, B_{r\theta}, C_{r\theta}$ are explicitly derived from the balance of convective stretching and relaxation for each individual species. Driven by the severe nonlinear growth of the scission rate with $\dot{\gamma}$, the stress contribution of species A experiences massive shear thinning beyond a critical deformation rate. When the solvent viscosity ratio $\beta_s$ is sufficiently small, this drastic structural breakdown exceeds the convective stress buildup. 
Consequently, the homogeneous viscometric flow curve exhibits a region with a strictly negative slope \cite{salmon2003velocity,lopez2004shear}, as captured in Fig.~\ref{flow_curve}.
\begin{figure}[htbp]
    \centering
    \begin{subfigure}{\linewidth}
        \centering
        \includegraphics[height=0.25\textheight]{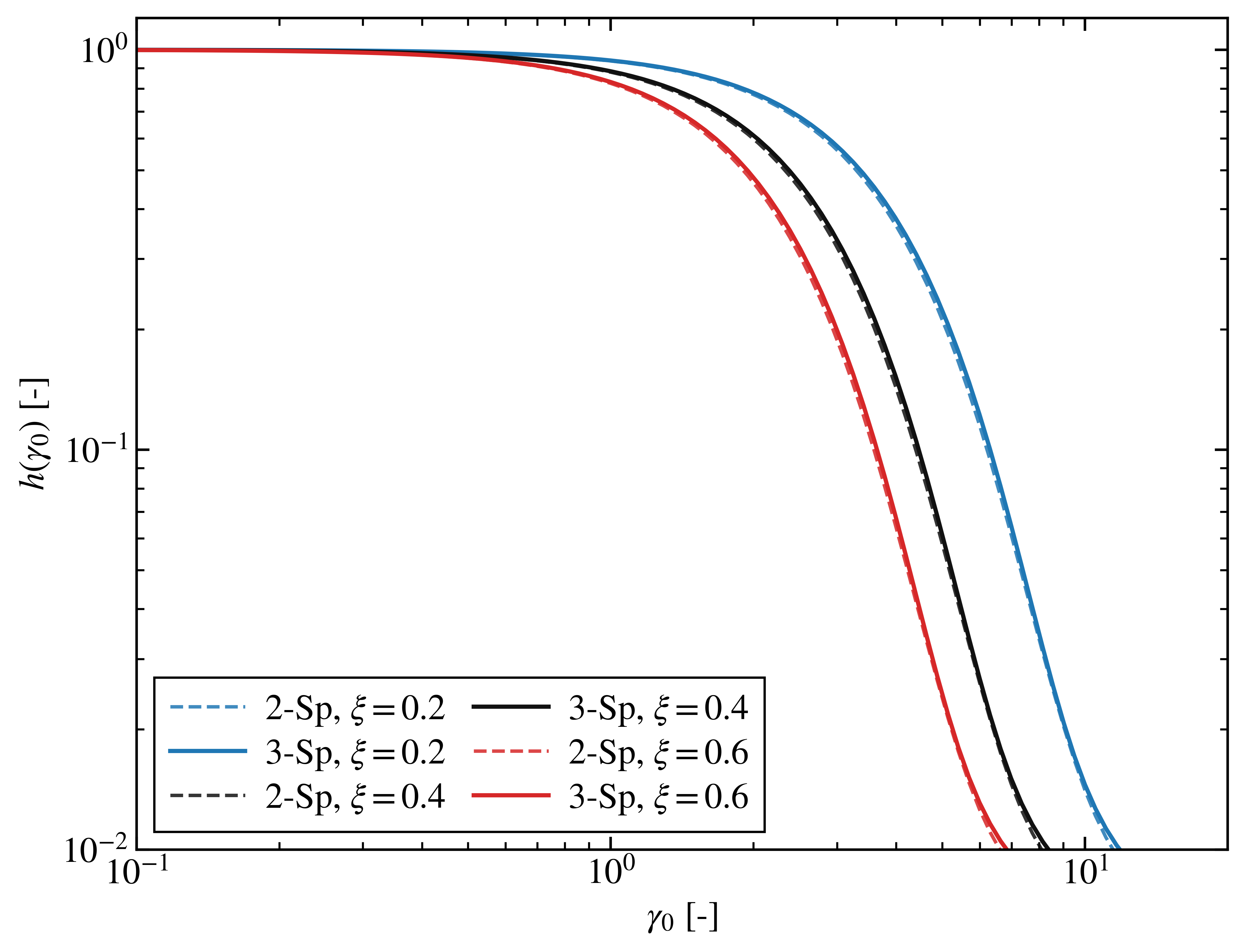}
        \caption{The nonlinear damping function $h(\gamma_0)$.}
    \end{subfigure}
    \begin{subfigure}{\linewidth}
        \centering
        \includegraphics[height=0.25\textheight]{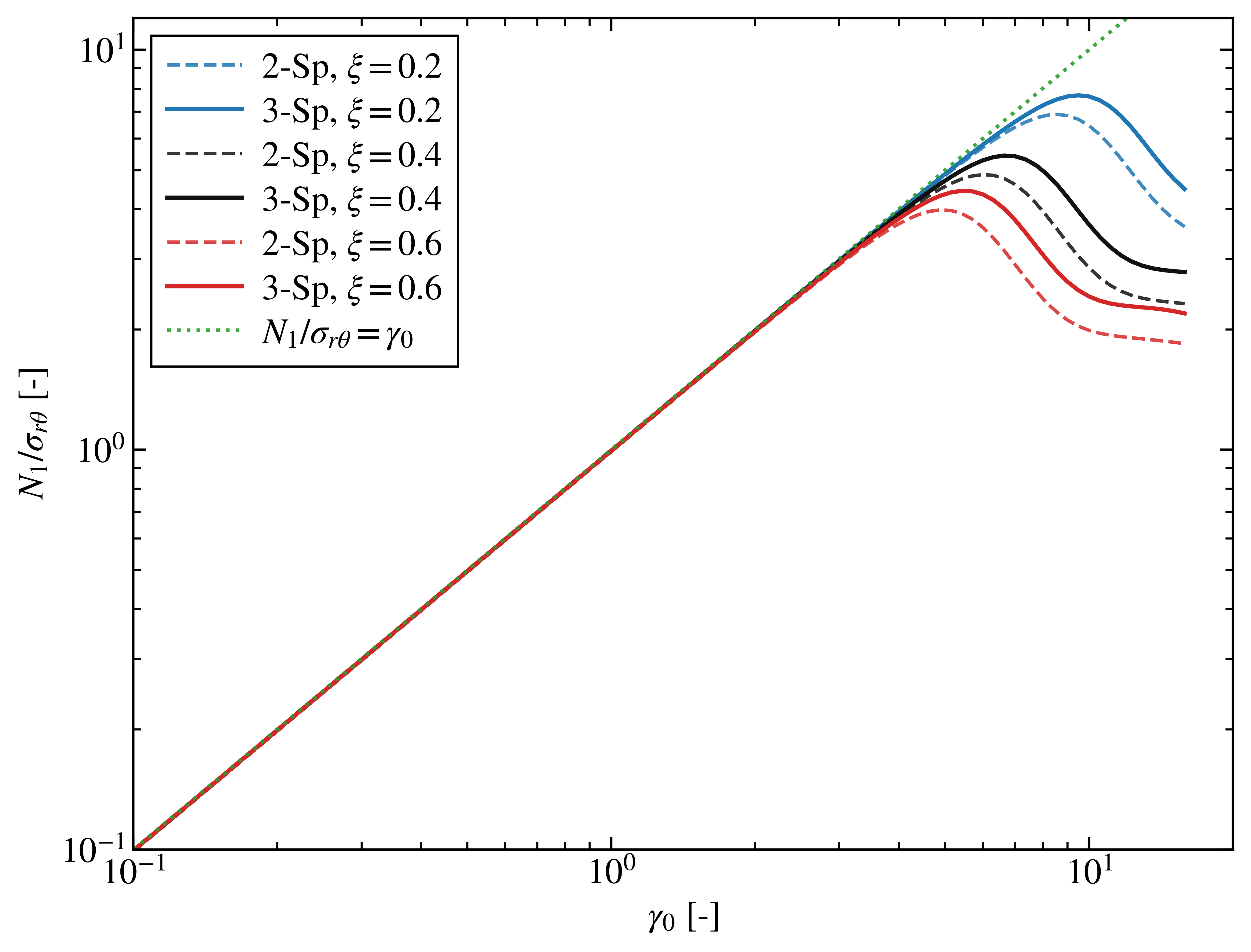}
        \caption{The Lodge-Meissner relation $N_1/\sigma_{r\theta}$.}
    \end{subfigure}

    \caption{Comparison of the nonlinear step strain responses between the classic 2 species and the proposed 3 species cascade models for varying non-affine slip parameters $\xi$.}
    \label{fig:damping_function}
\end{figure}

This non-monotonic homogeneous flow curve serves as the classical macroscopic signature of a constitutive instability. Within the shaded negative slope region in Fig.~\ref{flow_curve}, the assumed homogeneous flow field becomes mathematically unstable \cite{fielding2006nonlinear}. As a result, the fluid spontaneously separates into coexisting macroscopic layers of high and low shear rates. This structural transition constitutes the classical shear banding phenomenon.
Crucially, by independently regulating the physical parameters within the three-species regimes, such as the intermediate relaxation time $\lambda_B$ or the short chains density $n_C^0$, we can precisely map the depth and width of this unstable viscometric region, thereby dictating the critical onset conditions for shear banding.

\subsection{Transient step strain and stress relaxation} 
We next turn from steady viscometric flow to the transient nonlinear response under a step shear strain. In this standard rheological test, a shear strain $\gamma_0$ is imposed instantaneously at $t=0$, and the system is then allowed to relax under quiescent conditions with $\dot{\gamma}=0$ for $t>0$. At the moment immediately after deformation, $t=0^+$, the gel-network is assumed to deform affinely. This produces an initial stress state proportional to the imposed strain, so that the model recovers the usual neo-Hookean elastic limit before any structural rearrangement or relaxation takes place. As time progresses, the large initial stretch begins to modify the structural integrity of the fluid. Within the partially extending strand convection framework, the strong initial elastic deformation leads to a rapid increase in the scission rate of species A. This structural breakdown causes the primary network to rupture into species B, which can then further break into the terminal species C. As a result, this rapid redistribution of the species at short times accelerates the decay of the accumulated viscoelastic stress compared with the quiescent thermodynamic relaxation. At the macroscopic level, this accelerated structural breakdown leads to a clear deviation from the linear viscoelastic envelope. To quantify this nonlinear departure, we evaluate the transient stress response through the nonlinear relaxation modulus, $G(t,\gamma_0)=\sigma_{r\theta}(t,\gamma_0)/\gamma_0$. At large strain amplitudes ($\gamma_0 \gg 1$), the enhanced scission mechanism leads to a pronounced strain softening response. This nonlinear reduction in material memory is commonly described by the classical damping function $h(\gamma_0)$, which isolates the strain dependent decay of the relaxation amplitude \cite{brown1997tests, miller2007transient,Vasquez2007}. As illustrated in Fig.~\ref{fig:damping_function}a, the model predicts a strong monotonic decrease of $h(\gamma_0)$, which is consistent with the marked structural disentanglement in highly strained wormlike micellar solutions. Beyond the relaxation of the shear stress alone, an important test for the constitutive model is its tensorial consistency under finite deformation. For step strain kinematics, the Lodge–Meissner relation requires that the ratio of the first normal stress difference to the shear stress be equal to the applied strain, namely $N_1/\sigma_{r\theta}=\gamma_0$. Since the instantaneous scission process preserves the tensorial proportionality of the conformation tensors during mass transfer between species, the proposed three-species model naturally satisfies this continuum mechanical identity. This is confirmed in Fig.~\ref{fig:damping_function}b, where the Lodge–Meissner relation remains valid throughout the step strain response. The result shows that the model can describe the nonlinear rupture process without generating unphysical kinematic behavior.

\section{Inhomogeneous flow predictions}\label{sec-inhomo}
The homogeneous analysis in Section~\ref{sec-homo} shows that the present three-species cascade breakage
model admits a non-monotone viscometric response over an intermediate range of shear rates.
Once the imposed deformation enters this regime, a spatially uniform shear state is no longer
sustained, and the flow reorganizes into coexisting regions with different local shear rates and
microstructural compositions \cite{porte1997inhomogeneous, salmon2003velocity, hu2005kinetics, lopez2004shear}. To describe this shear banding transition, we therefore turn to
the full inhomogeneous problem.

We consider a cylindrical Couette geometry, in which the fluid is confined between two
concentric cylinders, with the inner cylinder rotating and the outer cylinder fixed. Denoting
the gap width by $d=R_o-R_i$, we characterize the imposed deformation by the Deborah
number $De=\lambda_{\mathrm{eff}}V_i/d$, where $V_i$ is the velocity of the inner cylinder.
Under the assumptions of axisymmetry and incompressibility, the flow satisfies
$\nabla\cdot\mathbf{v}=0$, and the velocity field reduces to the purely azimuthal form $\mathbf{v}=(0,v_\theta(r,t),0),$
so that all spatial variation occurs only in the radial direction. 

The governing equations are given by the one dimensional radial form of the coupled
number density, conformation, and momentum equations introduced in Section~\ref{sec-math}. As in
earlier inhomogeneous studies of wormlike micellar models, nonlocal diffusion is retained in
order to regularize the band interface and select the physically relevant banded state \cite{olmsted2000johnson, radulescu2003time,fielding2006nonlinear}.
The velocity satisfies no-slip conditions at the two cylinders, while the number densities and
conformation tensors satisfy zero flux boundary conditions. In the creeping flow limit, the
shear stress remains constrained by the Couette geometry, even though the velocity,
conformation, and concentration fields may become strongly heterogeneous across the gap.

Within this framework, the present three-species model allows us to examine not only the
macroscopic formation of low and high shear bands, but also the accompanying radial
redistribution of the three micellar populations. In particular, the intermediate species
introduces an additional structural transition layer between the gel-network dominated region and
the strongly degraded short chains region. In the following analysis, we first examine the steady banded states under shear rate control. We then study the transient start-up dynamics that lead to these states, and finally the multistep relaxation after shear cessation.

\subsection{Steady response}\label{subsec-inhomo-steady}
\begin{figure}[htbp]
    \centering
    \includegraphics[height=0.25\textheight]{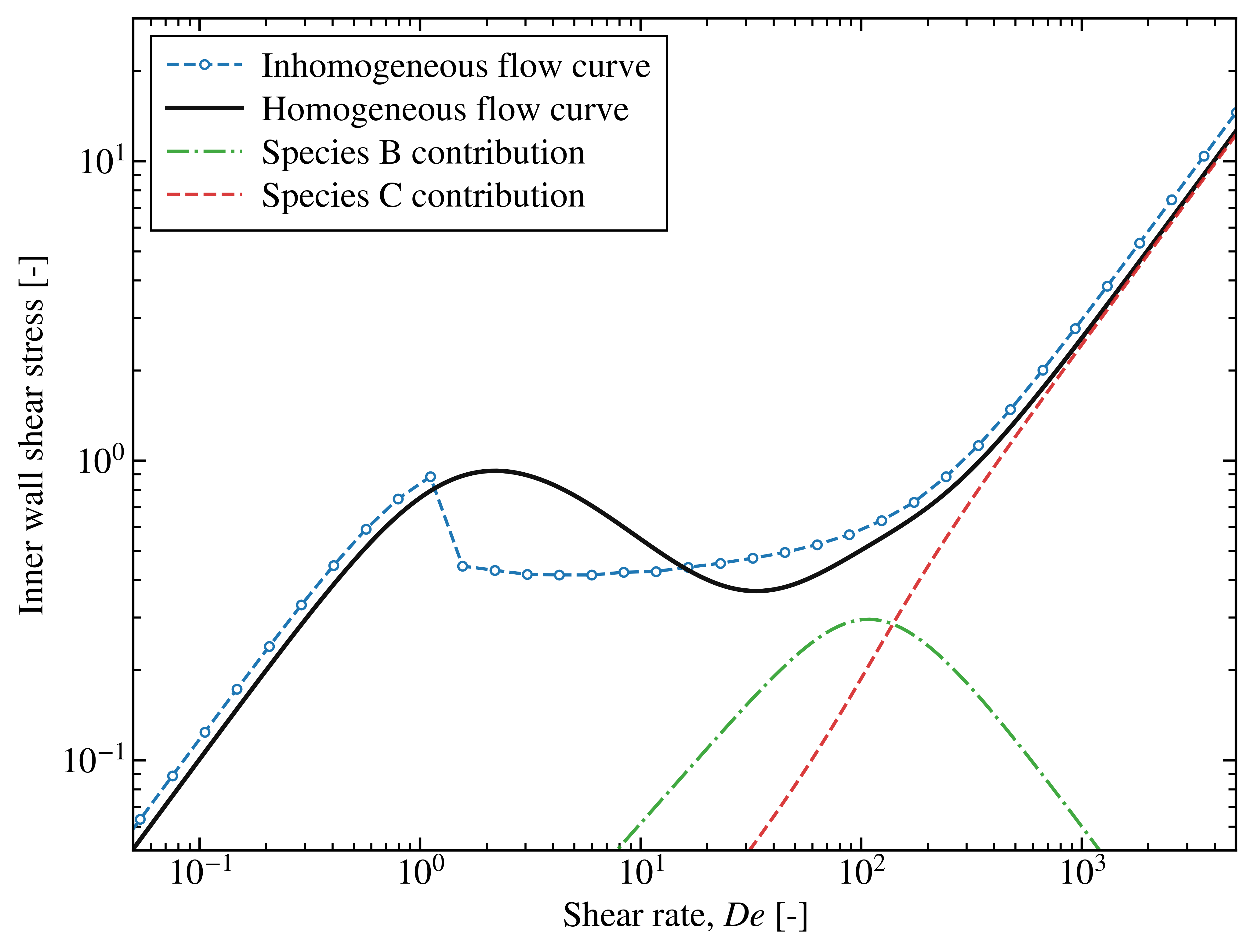}
    \caption{Steady wall shear stress as a function of the Deborah number under shear rate control, where the plateau in the inhomogeneous flow curve indicates the coexistence of shear bands across the Couette gap.}
    \label{fig:steady_flow_curve}
\end{figure}

\begin{figure}[htbp]
    \centering
    \begin{subfigure}{\linewidth}
        \centering
        \includegraphics[height=0.25\textheight]{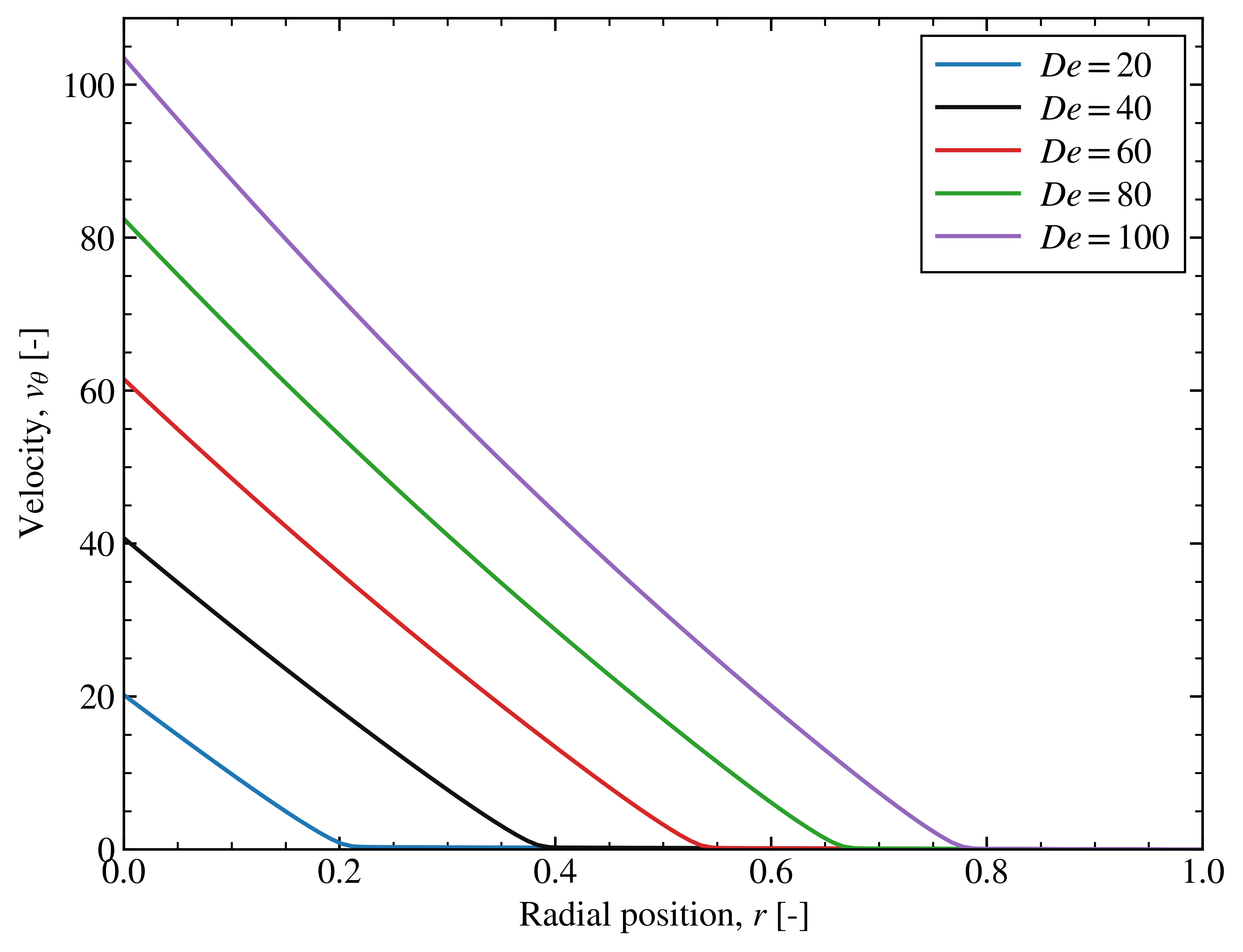}
        \caption{Radial velocity profiles $v_\theta(r)$ across the Couette gap.}
    \end{subfigure}
    \hfill
    \begin{subfigure}{\linewidth}
        \centering
        \includegraphics[height=0.25\textheight]{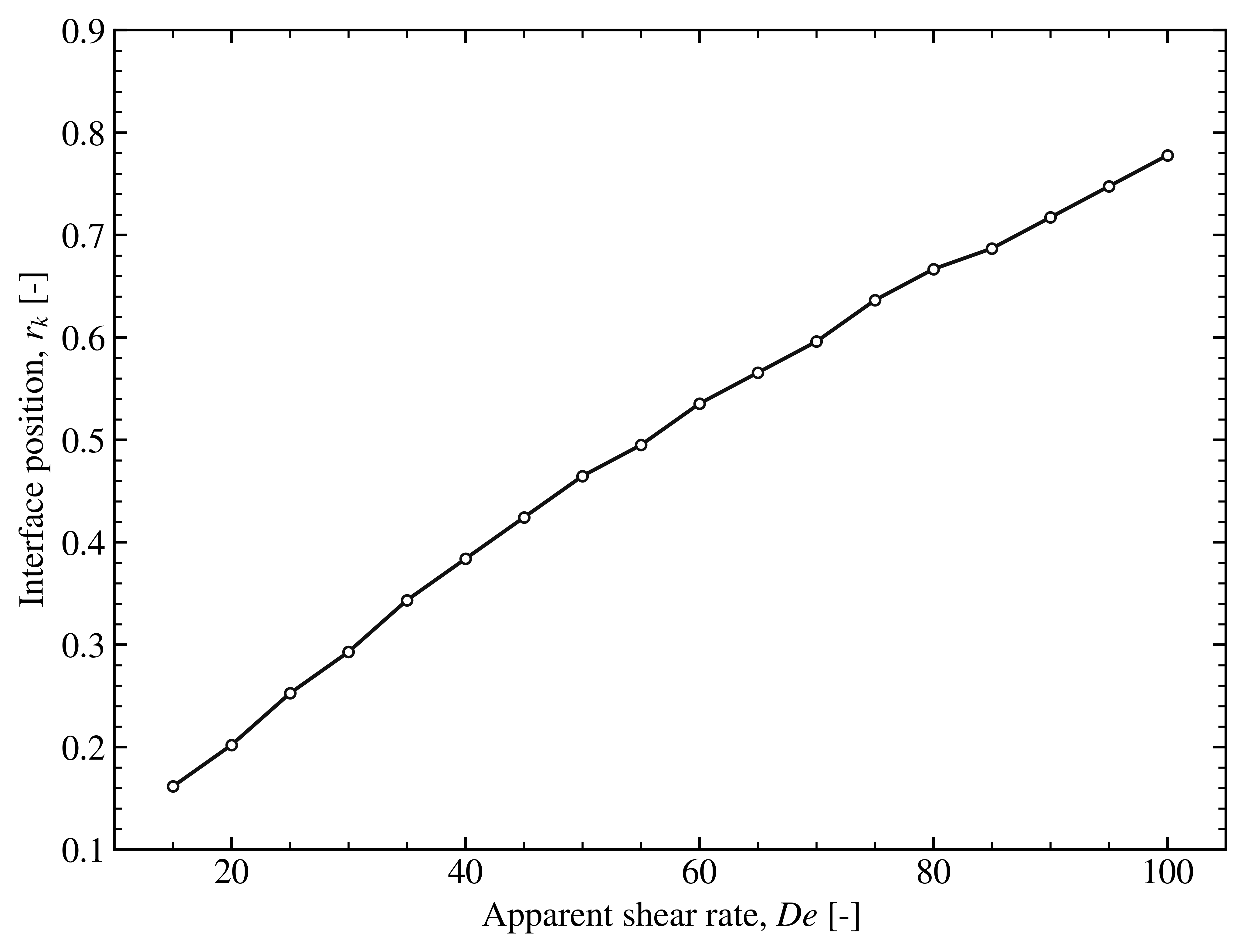}
        \caption{Evolution of the shear band interface (kink position, $r_k$) as a function of the applied apparent shear rate.}
    \end{subfigure}

    \caption{Steady state shear banding kinematics of the three-species cascade VCM model at high apparent shear rates ($De \ge 20$).}
    \label{fig:steady_velocity_profiles}
\end{figure}

We begin with the steady states obtained under shear rate control. For each prescribed Deborah number $De$, the full inhomogeneous system is evolved until the velocity, number densities, and conformation tensors reach steady values. The resulting wall shear stress is plotted against $De$ in Fig.~\ref{fig:steady_flow_curve}. Consistent with the homogeneous analysis in Section~\ref{sec-homo}, the decreasing branch of the viscometric flow curve is not realized in the inhomogeneous calculation. Instead, it is replaced by a plateau region, which indicates the coexistence of two shear bands across the gap \cite{salmon2003velocity, hu2005kinetics, miller2007transient}.
The corresponding steady velocity profiles are shown in
Fig.~\ref{fig:steady_velocity_profiles}. For small $De$, the profile is close to linear, which
suggests that the material remains in a single low shear state. As $De$ enters the plateau
region, the profile develops a visible kink. This kink separates a high shear band near the
inner rotating cylinder from a low shear band near the outer wall. With further increase of
$De$, the interface moves gradually outward, and the high shear band occupies a larger part
of the gap \cite{salmon2003velocity, hu2005kinetics, Zhou2014}. When $De$ exceeds the plateau region, the flow returns to a single band state.

\begin{figure}[htbp]
    \centering
    \begin{subfigure}{\linewidth}
        \centering
        \includegraphics[height=0.25\textheight]{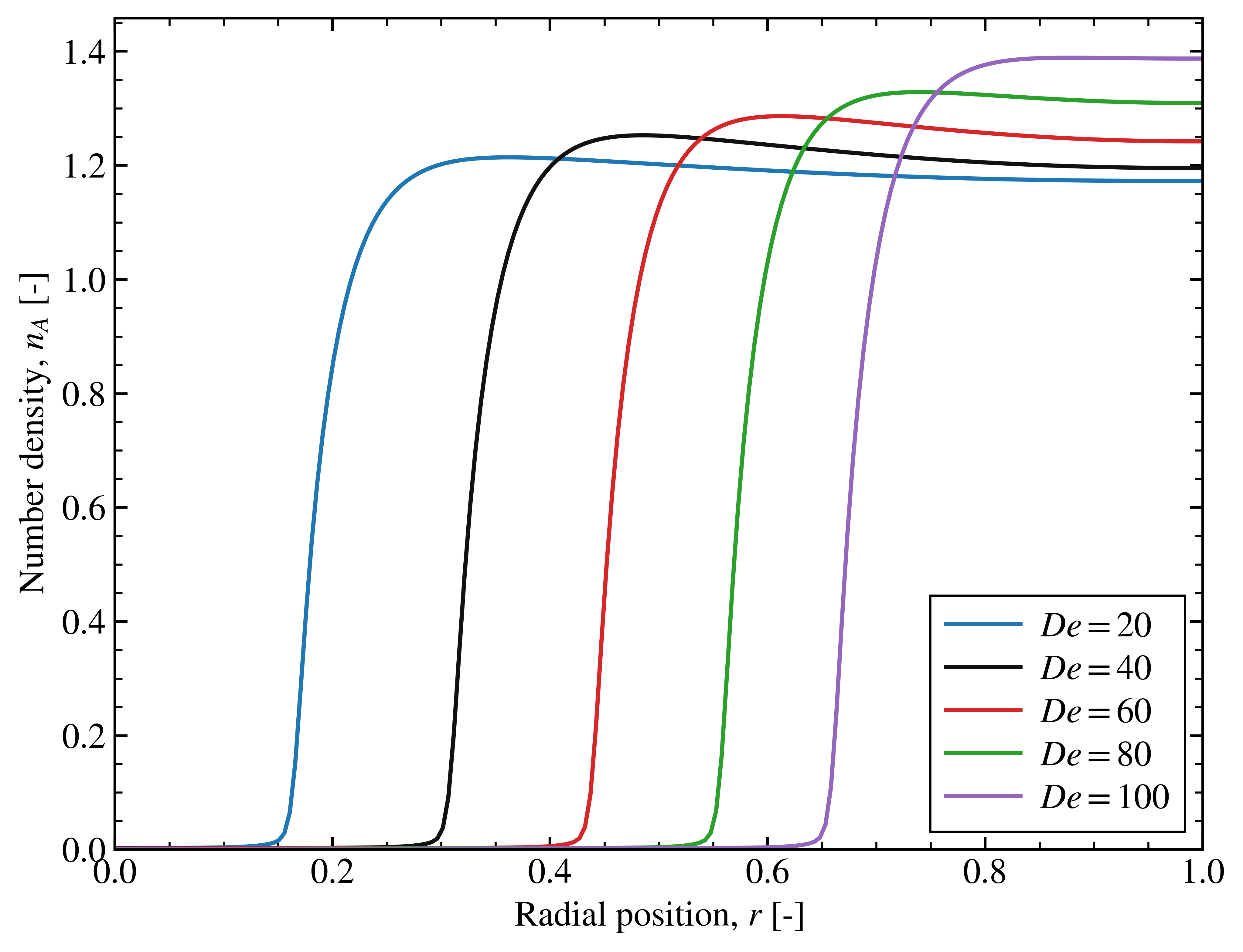}
        \caption{Number density of species $n_A$.}
    \end{subfigure}
    \hfill
    \begin{subfigure}{\linewidth}
        \centering
        \includegraphics[height=0.25\textheight]{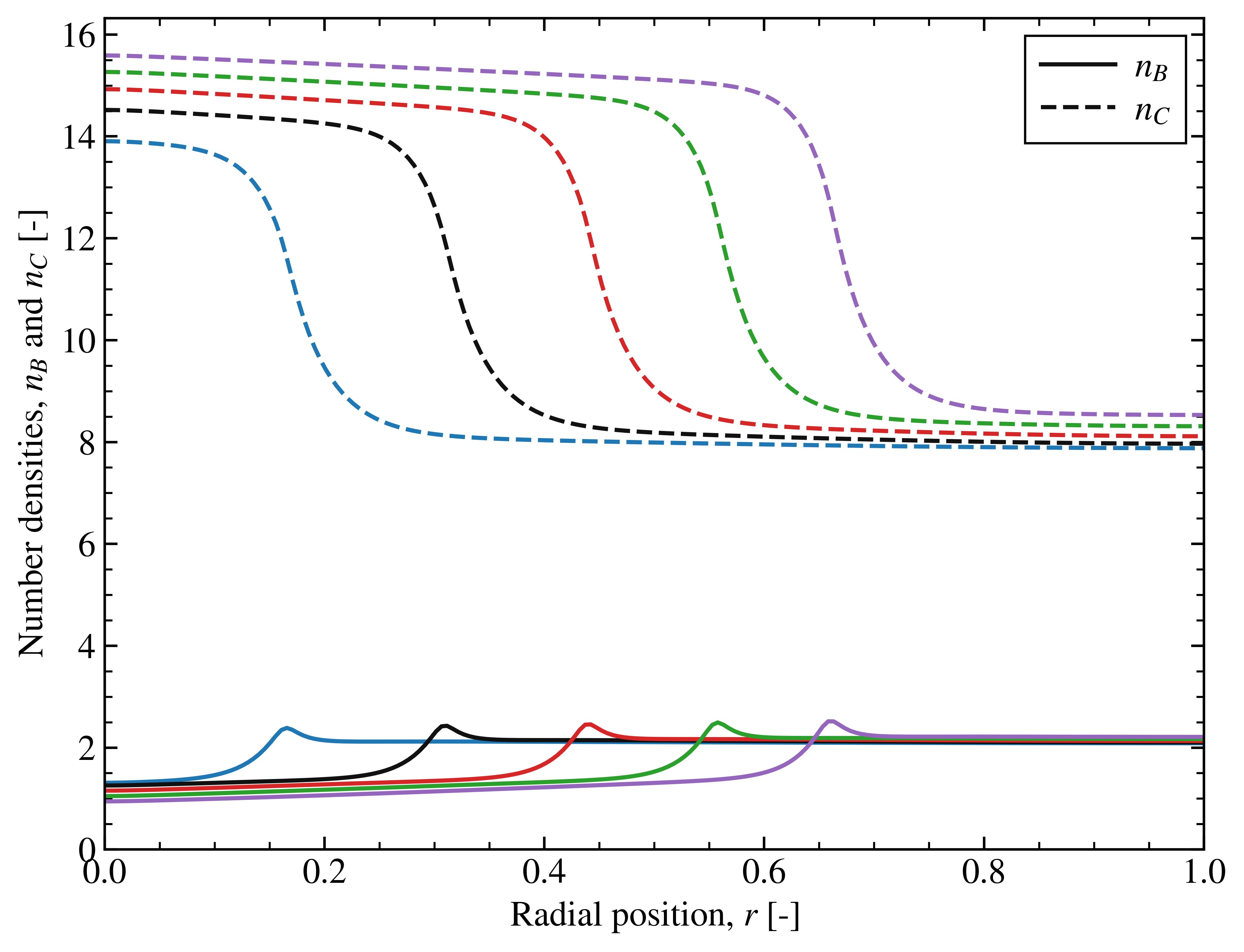}
        \caption{Number densities of species $n_B$ and $n_C$.}
    \end{subfigure}

    \caption{Steady state radial number density profiles evaluated at high apparent shear rates ($De = 20, 40, 60, 80, 100$).}
    \label{fig:steady_concentration_profiles}
\end{figure}

\begin{figure}[htbp]
    \centering
    \begin{subfigure}{\linewidth}
        \centering
        \includegraphics[height=0.25\textheight]{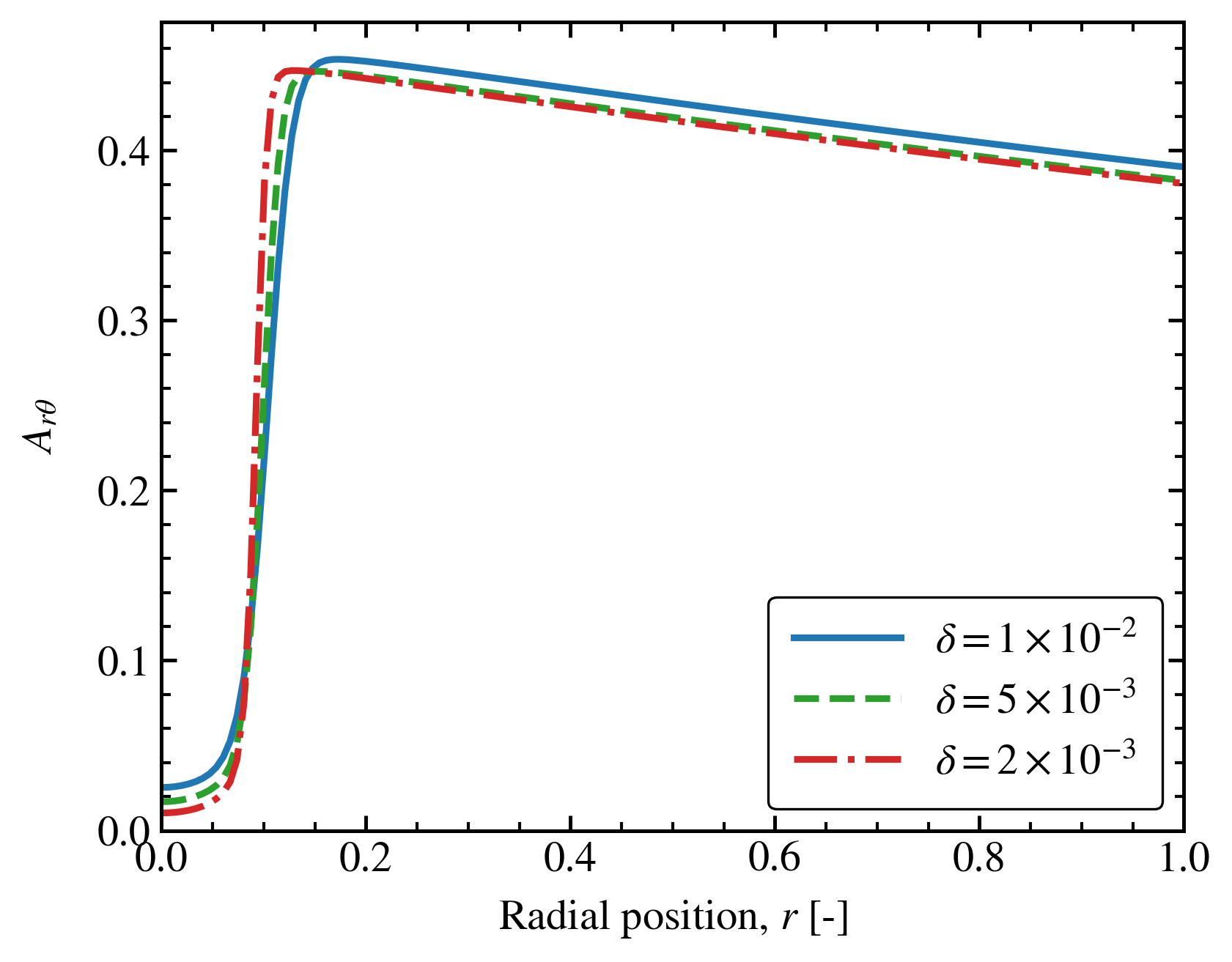}
        \caption{Shear stress contribution from species A $A_{r\theta}$.}
    \end{subfigure}
    \begin{subfigure}{\linewidth}
        \centering
        \includegraphics[height=0.25\textheight]{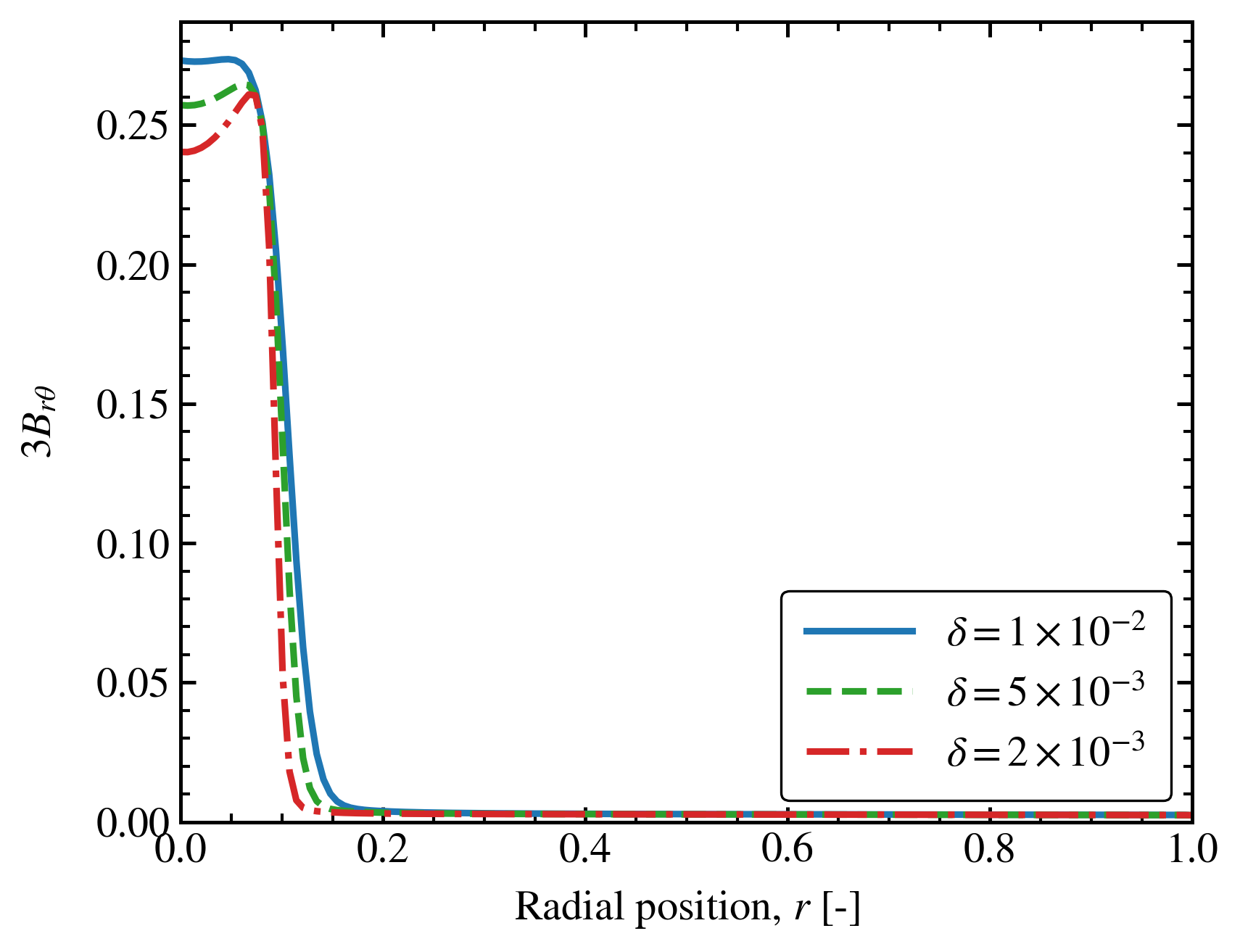}
        \caption{Shear stress contribution from species B $3B_{r\theta}$.}
    \end{subfigure}
    \begin{subfigure}{\linewidth}
        \centering
        \includegraphics[height=0.25\textheight]{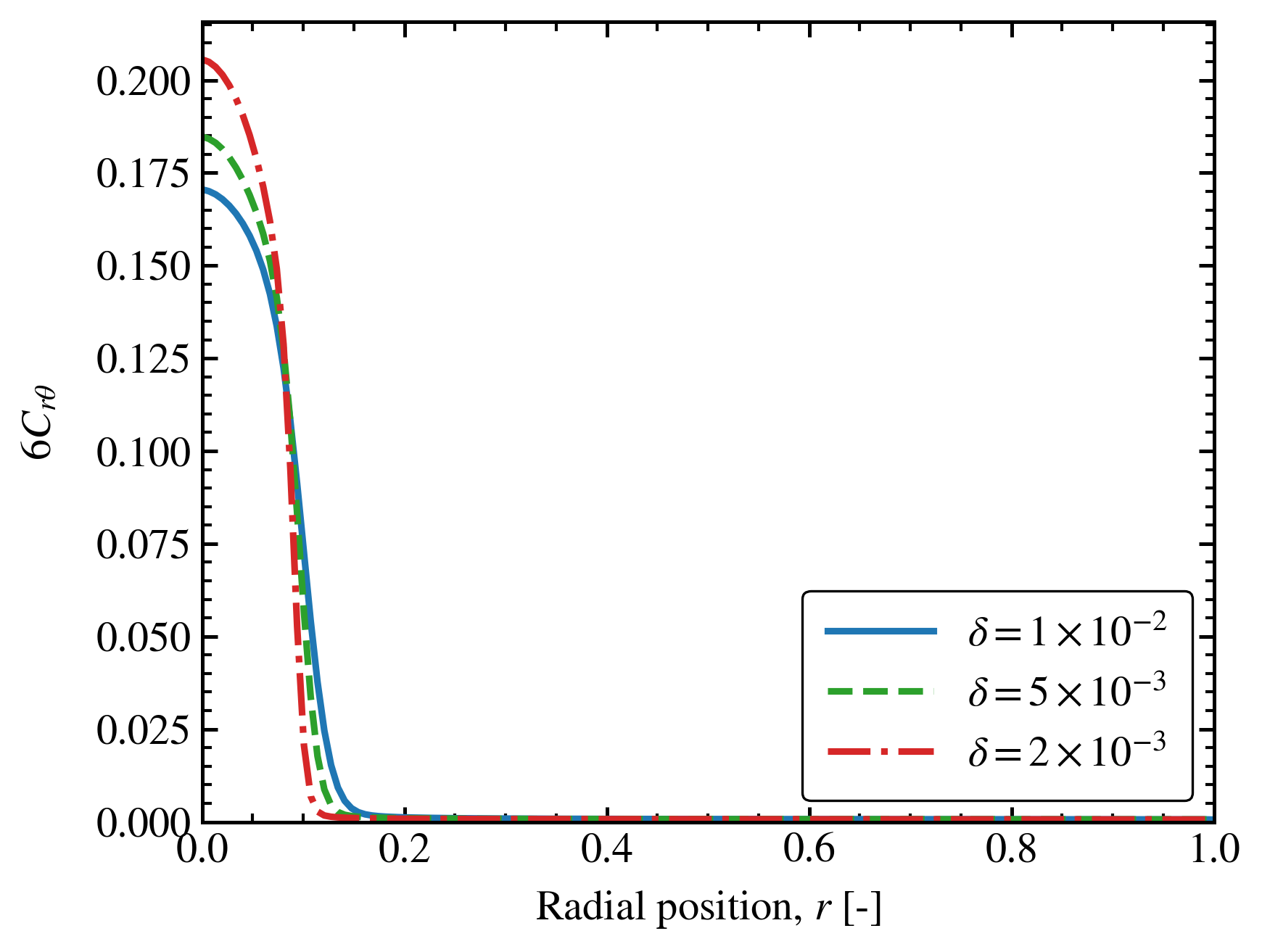}
        \caption{Shear stress contribution from species C $6C_{r\theta}$.}
    \end{subfigure}
    \caption{Spatial profiles of the decoupled shear stress contributions across the Couette gap for selected values of the diffusion parameter.}
    \label{fig:steady_stress_decomposition}
\end{figure}

To understand the structural change behind this kinematic transition, we next inspect the
steady radial distributions of the three species. Fig.~\ref{fig:steady_concentration_profiles}
shows that the three-species model produces a clear spatial partition of the micellar
populations. In the low shear band, species A remains dominant, indicating
that the local microstructure is only weakly degraded. In contrast, in the high shear band
near the inner wall, the stronger local deformation promotes the cascade breakage process and
reduces the population of the slow, highly elastic species. At the same time, the shorter and
faster relaxing species become more important. Between these two regions, species B forms a transition layer, linking the intact gel-network dominated band and the strongly
degraded band. The presence of this intermediate layer allows the model to resolve the structural transition across the band interface in a way that is not available in a simpler two-species description.
The same trend can also be seen in the stress decomposition. In
Fig.~\ref{fig:steady_stress_decomposition}, the total shear stress is separated into the
contributions carried by the different species. In the low shear band, the main contribution
comes from the gel-network. In the high shear band, this contribution
decreases, while the stress supported by the shorter species becomes relatively more important.
The intermediate species provides a gradual transfer between these two limiting regimes. In
this way, the selected steady plateau is associated not only with a jump in the local shear
rate, but also with a redistribution of stress support among the three structural levels.

These steady results show that shear banding in the present model involves both kinematic localization and structural partition across the gap. The stress plateau and kinked velocity profiles reflect the macroscopic banded state, while the radial redistribution of the three species reveals its underlying microstructural basis.

\subsection{Transient response}

We now turn to the transient evolution toward the steady banded states described above.
For each prescribed Deborah number $De$, the system is started from the equilibrium state
and then subjected to shear by imposing the wall motion at $t=0$. We track the time
evolution of the shear stress, velocity field, and species distributions in order to examine
how the final inhomogeneous state is formed.

\begin{figure}[htbp]
    \centering
    \includegraphics[height=0.25\textheight]{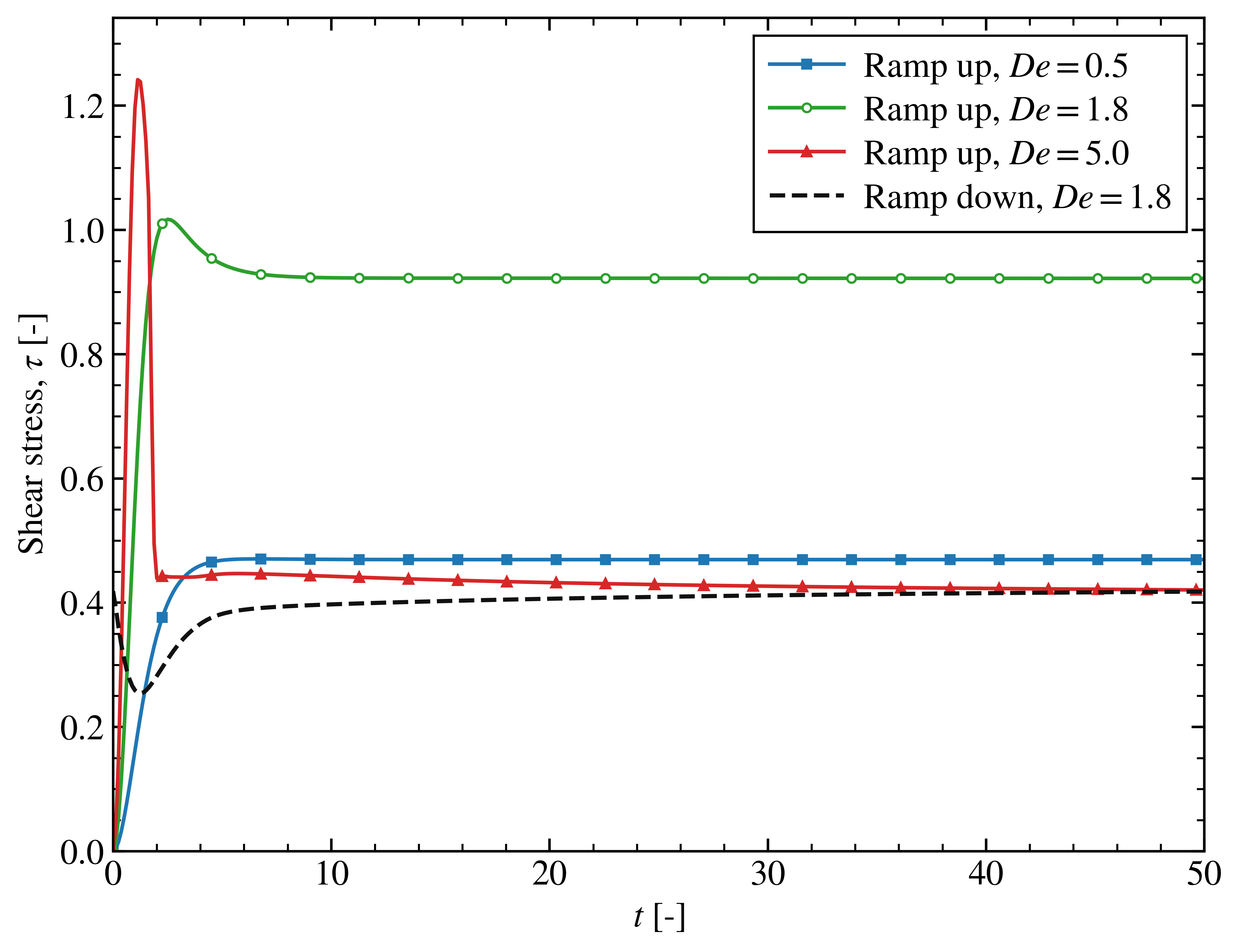}
    \caption{Time evolution of the macroscopic shear stress for the three-species model. The plot contrasts start-up flows from rest at various $De$ with a step-down flow to $De = 1.8$, highlighting both flow hysteresis and thixotropic stress recovery dynamics.}
    \label{fig:transient_stress_response}
\end{figure}

The transient shear stress response under both start-up and step-down flow protocols is shown in Fig.~\ref{fig:transient_stress_response}. During start-up from rest, for small apparent shear rates ($De=0.5$), the stress rises smoothly from its initial value and approaches a homogeneous steady state without a pronounced stress overshoot. In contrast, when $De$ lies in or above the shear banding regime ($De=5.0$), the stress typically exhibits a strong overshoot before relaxing toward a much lower long-time value, indicating severe flow softening. This start-up behavior reflects the competition between elastic loading and microstructural breakdown. At early times, the gel-network is stretched by the imposed shear, supporting a rapidly increasing viscoelastic stress. As the deformation continues, the cascade breakage process accelerates, reducing the population of the slow elastic structures and leading to a stress drop from the overshoot toward the final banded state. To further probe the flow hysteresis and microstructural rebuilding dynamics, a step-down rate test is simulated by suddenly decreasing the shear rate from a fully degraded, high shear state $De=10.0$ to the metastable regime $De=1.8$. Unlike a simple monotonic decay, the transient stress exhibits a classical U-shaped thixotropic recovery. The initial abrupt stress drop corresponds to the instantaneous reduction in solvent viscous drag. The subsequent gradual stress increase captures the thixotropic rebuilding of the micellar network. Crucially, compared to reduced binary descriptions, the present three-species model naturally resolves the prolonged recovery timescale, as the terminal short fragments must undergo a hierarchical recombination process to rebuild the entanglements. Such macroscopic stress undershoots mirror the finite structural build-up time widely documented in transient step-down experiments and phenomenological models of wormlike micellar solutions \cite{bautista1999understanding, mewis2009thixotropy}. Ultimately, the stark divergence in the final steady state stress at $De=1.8$, depending on whether the system was started from rest or stepped down from a high shear rate, clearly demonstrates flow hysteresis, a hallmark of the shear banding transition.

\begin{figure}[htbp]
    \centering
    \includegraphics[height=0.25\textheight]{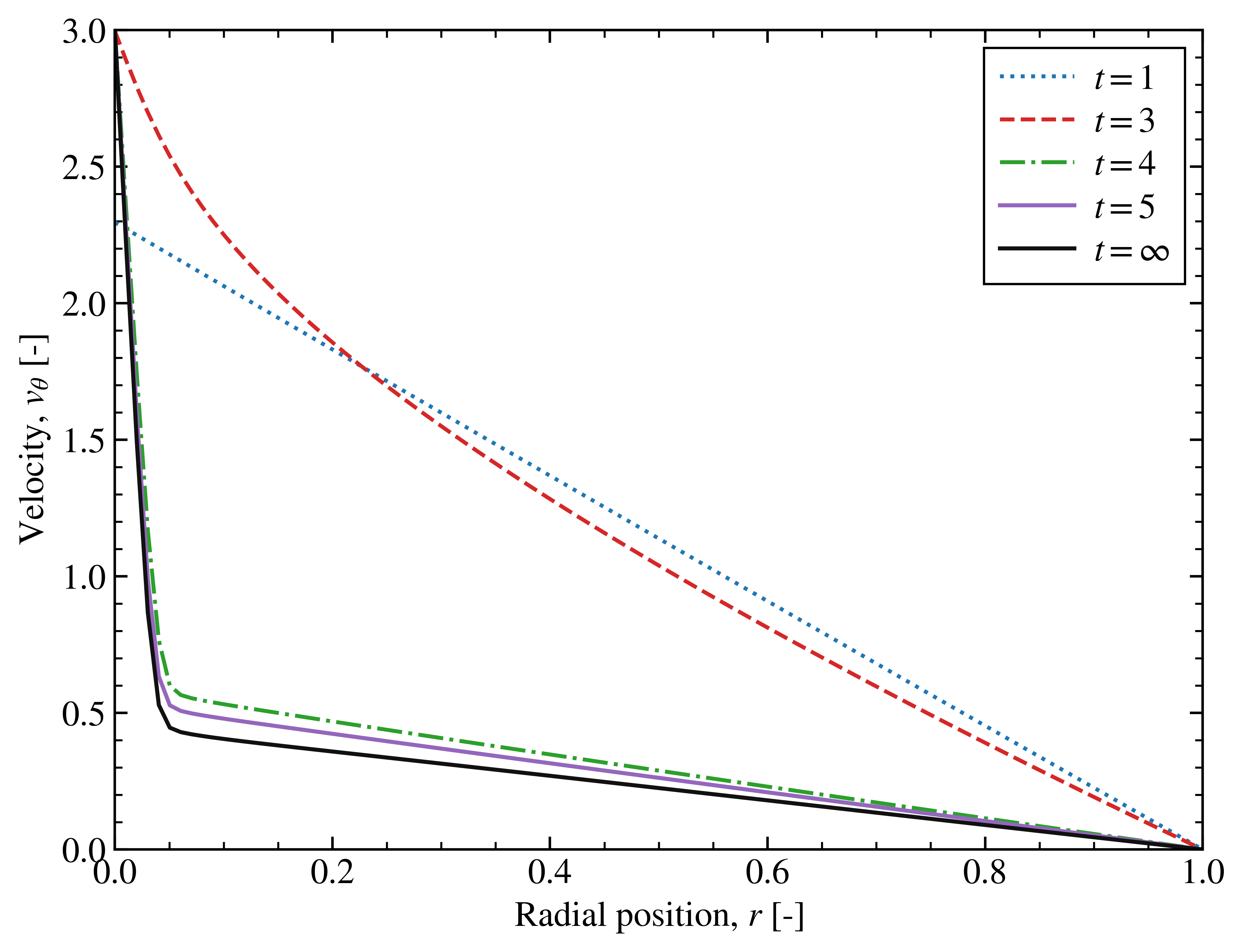}
    \caption{Transient velocity profiles at $De=3.0$ demonstrating the formation of a macroscopic shear band.}
    \label{fig:transient_velocity_evolution}
\end{figure}

\begin{figure}[htbp]
    \centering
    \begin{subfigure}{\linewidth}
        \centering
        \includegraphics[height=0.25\textheight]{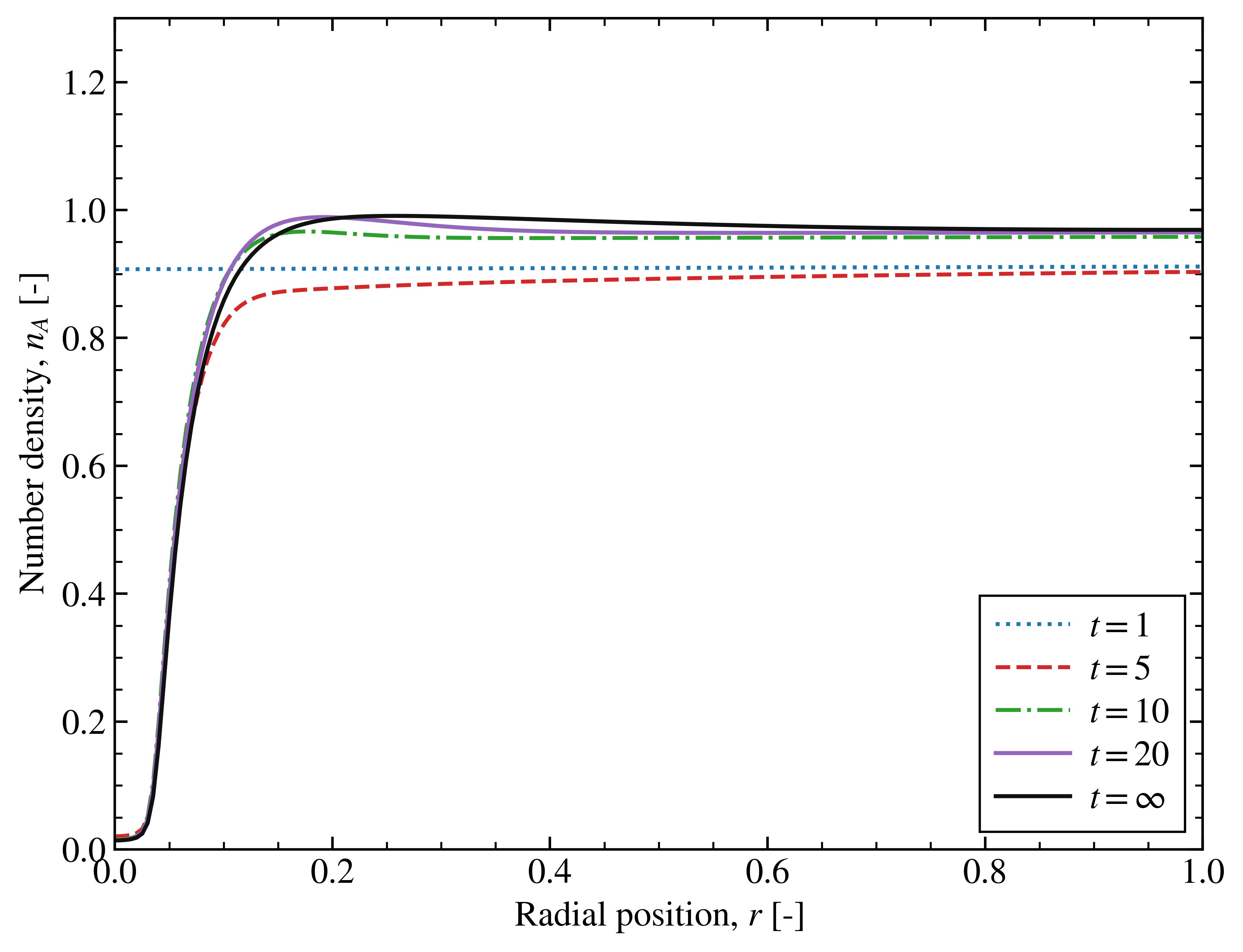}
        \caption{Transient concentration of species A, $n_A$.}
    \end{subfigure}
    \begin{subfigure}{\linewidth}
        \centering
        \includegraphics[height=0.25\textheight]{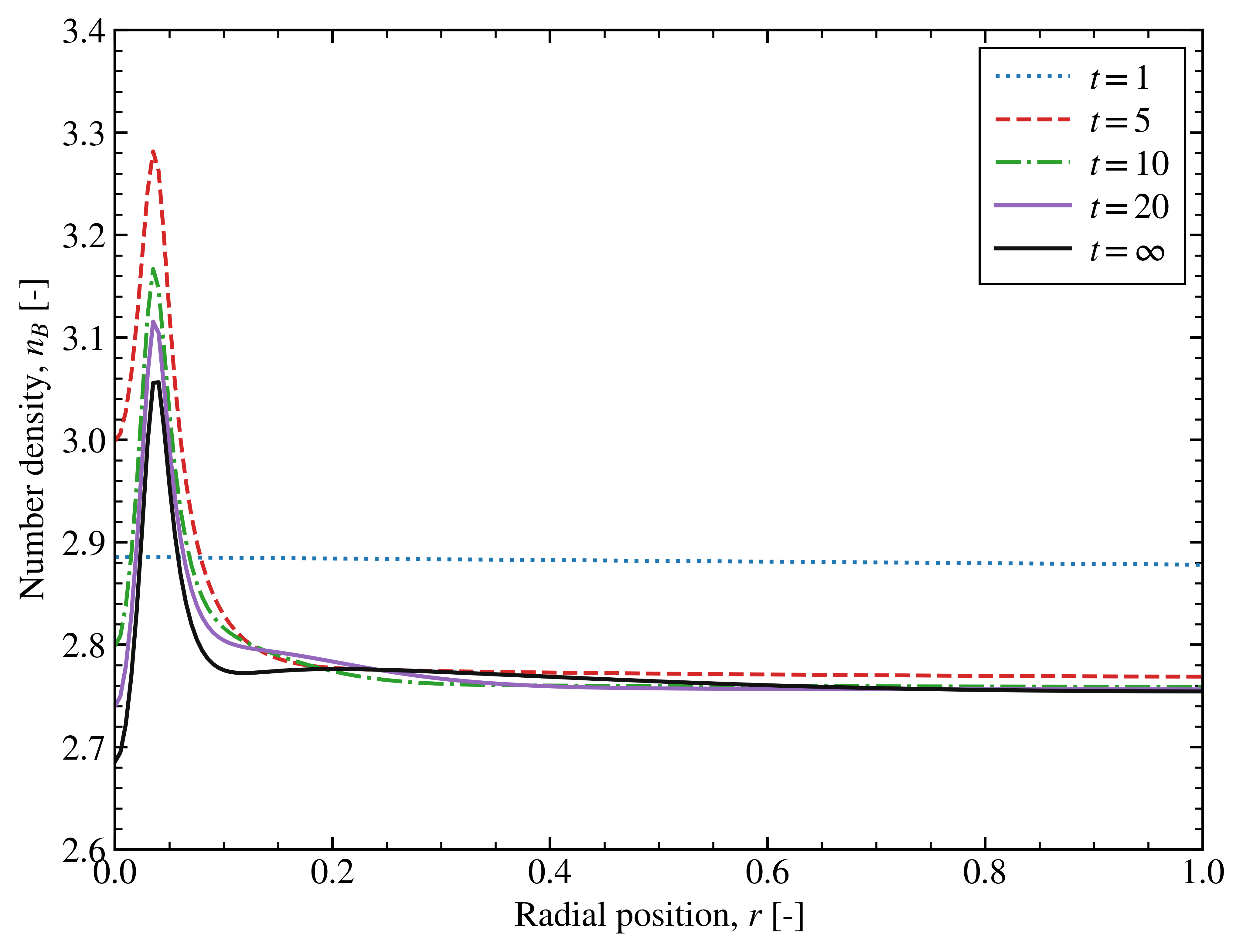}
        \caption{Transient concentration of species B, $n_B$.}
    \end{subfigure}
    \begin{subfigure}{\linewidth}
        \centering
        \includegraphics[height=0.25\textheight]{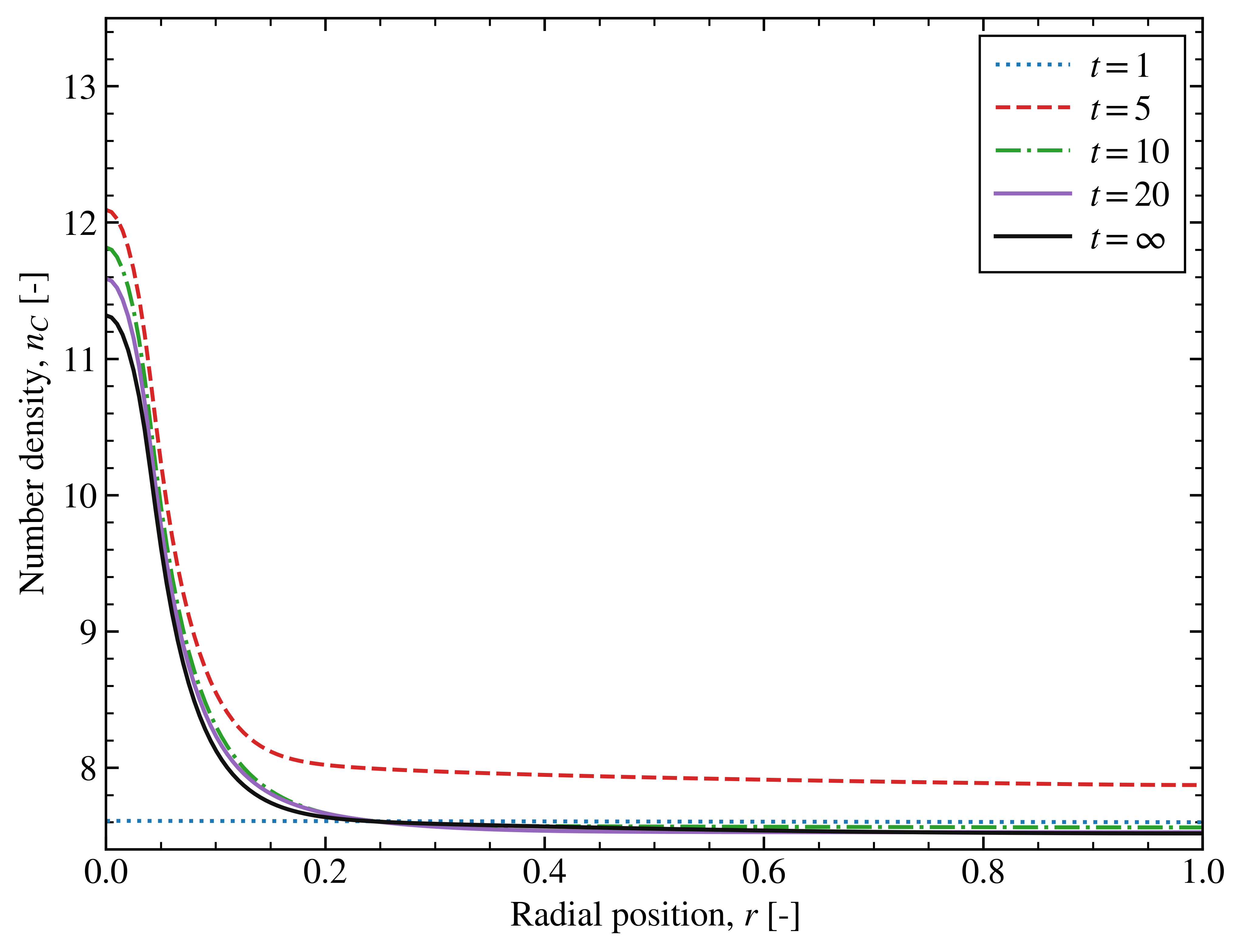}
        \caption{Transient concentration of species C, $n_C$.}
    \end{subfigure}
    \caption{Transient concentration profiles of the three species across the Couette gap during the formation of a macroscopic shear band.}
    \label{fig:transient_concentration_evolution}
\end{figure}

\begin{figure}[htbp]
    \centering
    \includegraphics[height=0.25\textheight]{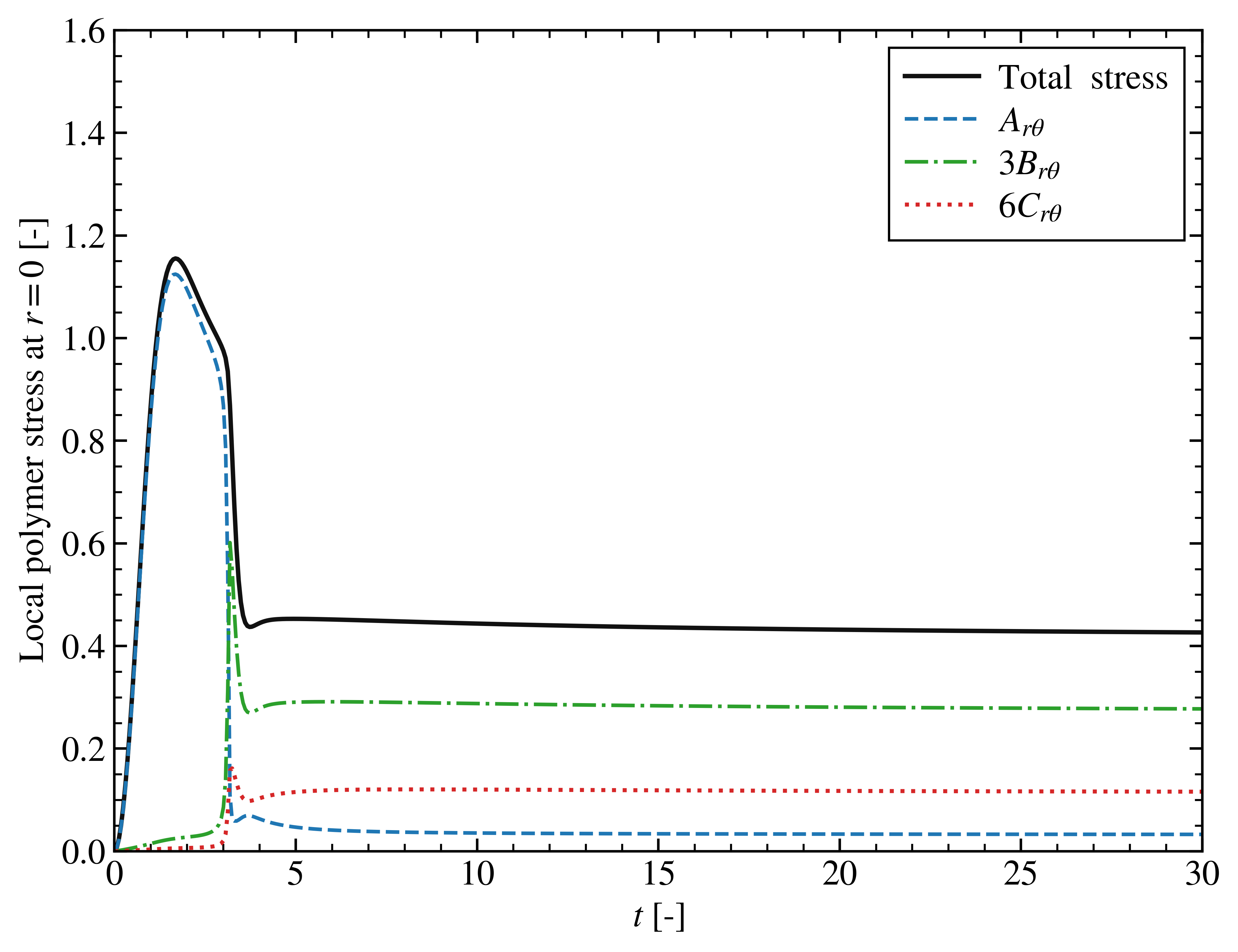}
    \caption{Transient stress partition at $r=0$ during the onset of shear banding ($De=3.0$). The total polymer stress is shown alongside the microstructural contributions from the gel-network $A_{r\theta}$, long chains $3B_{r\theta}$, and short chains $6C_{r\theta}$, demonstrating the microscopic stress transfer driven by cascade breakage.}
    \label{fig:transient_stress_partition}
\end{figure}

The evolution of the velocity profile is presented in
Fig.~\ref{fig:transient_velocity_evolution}. Shortly after the start of shearing, the profile remains close to that of a nearly homogeneous shear flow. As time increases, a localized region with
larger shear rate develops near the inner rotating cylinder. This localized region gradually
sharpens and evolves into a clear band interface, which then moves across the gap until the
steady two-band configuration is established \cite{miller2007transient, salmon2003velocity, Zhou2014}. Thus, the steady kinked profiles are not formed instantaneously, but emerge through a continuous reorganization
of the local shear field.
The transient microstructural evolution is shown in
Fig.~\ref{fig:transient_concentration_evolution}. At early times, the concentrations of the three
species remain close to their initial values, since the deformation has not yet produced a strong
spatial differentiation. Once the local stress becomes sufficiently large near the inner wall, the
breakage cascade is accelerated in that region. As a result, the concentration of the slow,
highly elastic species decreases, while the shorter and faster relaxing species increase. The
intermediate species again plays a transitional role: its concentration first grows in the region
where the long structures are being depleted, and then adjusts as the short species dominated
band is formed. In this sense, the transient process is not only a kinematic localization of the
velocity field, but also a gradual redistribution of mass among the three structural levels.
The corresponding transient stress partition is shown in
Fig.~\ref{fig:transient_stress_partition}. During the initial stage, the total stress is mainly carried
by the gel-network component. Near the overshoot, this contribution reaches its
maximum. At later times, as the local cascade breakage develops, the stress carried by this
component decreases, and the contributions from the intermediate and shorter species become
more visible. This transfer of stress support is consistent with the concentration changes
described above and provides a direct structural interpretation of the transient stress relaxation
after the overshoot.

Overall, the transient dynamics show that the approach to the steady banded state proceeds
through a sequence of physically clear stages: initial elastic loading, stress overshoot, spatial
localization of the shear field, and gradual redistribution of the three species across the gap.
Compared with a reduced two-species description, the present model resolves a broader structural
transition during this process, since the intermediate species introduces an additional time scale
and an additional channel for stress transfer. These features become even more evident after cessation of shear, where the model predicts a multistep relaxation process governed by the sequential recovery of the three micellar populations.

\subsection{Multistep shear relaxation}

We finally consider the relaxation process after cessation of shear. In this part, the system is
first driven under an imposed shear until a prescribed reference state is reached, and the wall
motion is then stopped at $t=t_s$. The subsequent evolution is governed entirely by the
internal stress relaxation, species interconversion, and diffusive redistribution of the
microstructure \cite{Zhou2014}. Our main interest here is to examine whether the present three-species model
produces a simple single stage decay or a more complex relaxation process.

\begin{figure}[htbp]
    \centering
    \includegraphics[height=0.25\textheight]{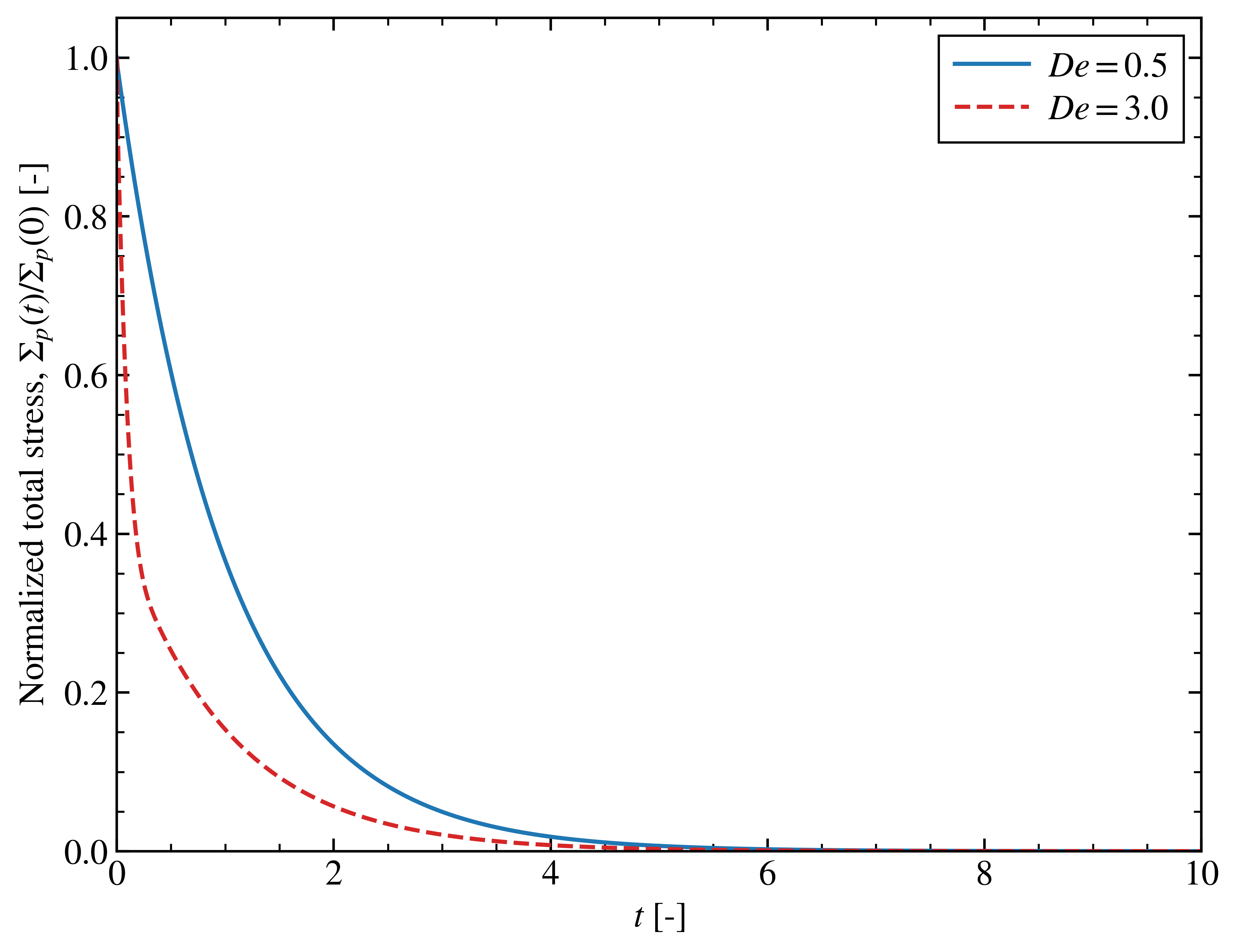}
    \caption{Stress relaxation profiles upon cessation of steady shear for $De=0.5$ and $De=3.0$. The multistep decay at the higher $De$ highlights the complex microstructural redistribution among the three species with highly separated relaxation times.}
    \label{fig:relaxation_stress_decay}
\end{figure}

The relaxation of the total shear stress is shown in
Fig.~\ref{fig:relaxation_stress_decay}. For weakly deformed states, the stress decreases
smoothly and approaches zero in a relatively simple manner. However, for states prepared in
the banding regime, the decay is no longer well described by a single characteristic time. The
stress typically exhibits a fast initial drop, followed by one or more slower stages before the
system fully returns to equilibrium. This behavior indicates that the relaxation process is
distributed over several structural levels rather than being controlled by a single dominant mode.

\begin{figure}[htbp]
    \centering
    \includegraphics[height=0.25\textheight]{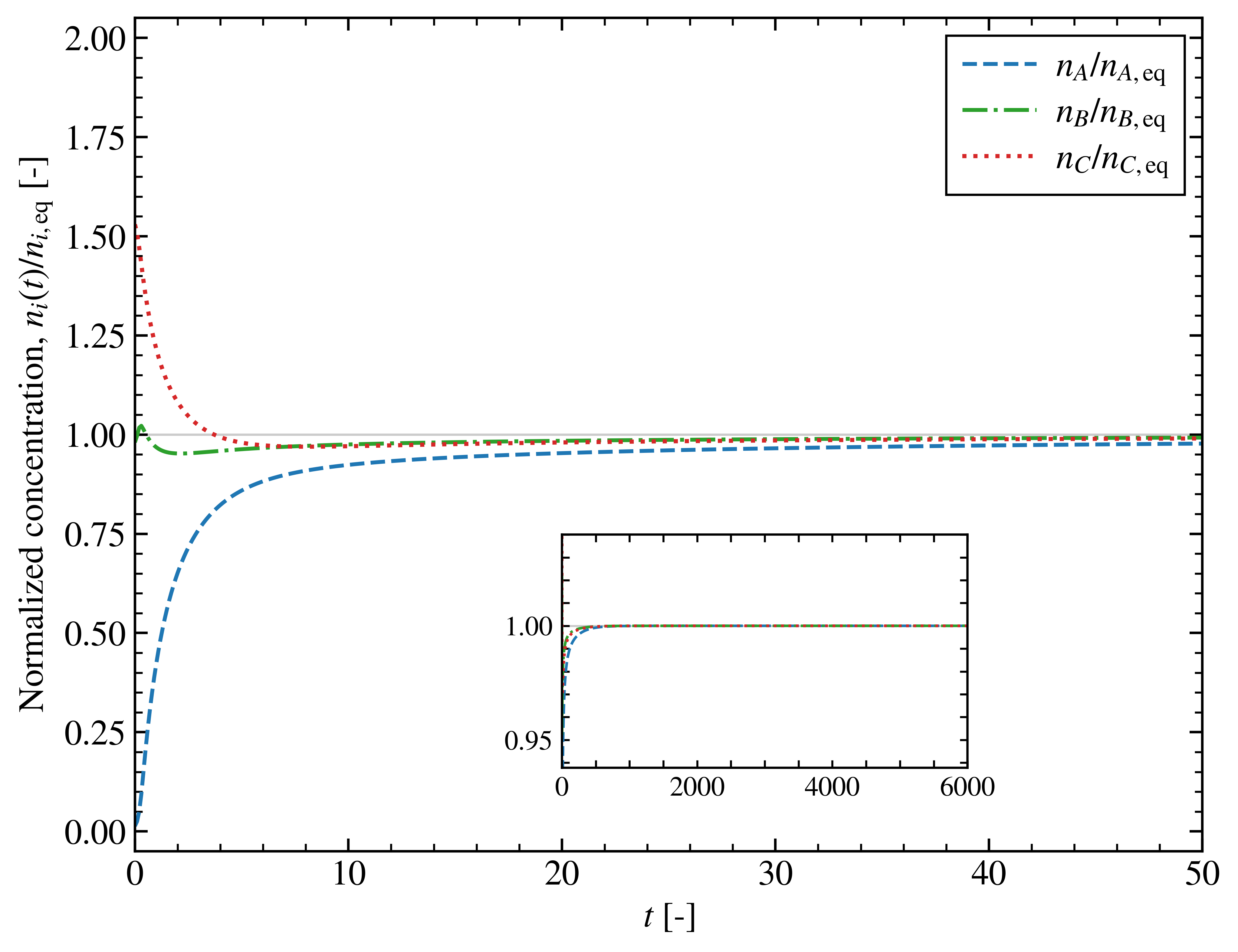}
    \caption{Microstructural relaxation after shear cessation ($De=3.0$). The main figure shows the rapid localized species interconversion ($t \le 50$), while the inset captures the long-time spatial diffusion required to fully restore the homogeneous global equilibrium ($t \to 6000$).}
    \label{fig:relaxation_concentration_evolution}
\end{figure}

To better understand this multistep decay, we inspect the time evolution of the species
concentrations after shear cessation. As shown in
Fig.~\ref{fig:relaxation_concentration_evolution}, the shortest and fastest relaxing species C responds first. Its associated stress relaxes quickly once the external deformation is removed.
At the same time, the intermediate species B continues to evolve through the cascade exchange,
acting as a bridge between the rapidly relaxing short fragments and the slowly recovering
network-rich population. The gel-network species A relaxes on a later time scale, and its
recovery toward equilibrium is accompanied by a gradual rebuilding of the network structure.
As a result, the full relaxation process proceeds through a sequence of partially overlapping
stages rather than a single monotone adjustment.

\begin{figure}[htbp]
    \centering
    \includegraphics[height=0.25\textheight]{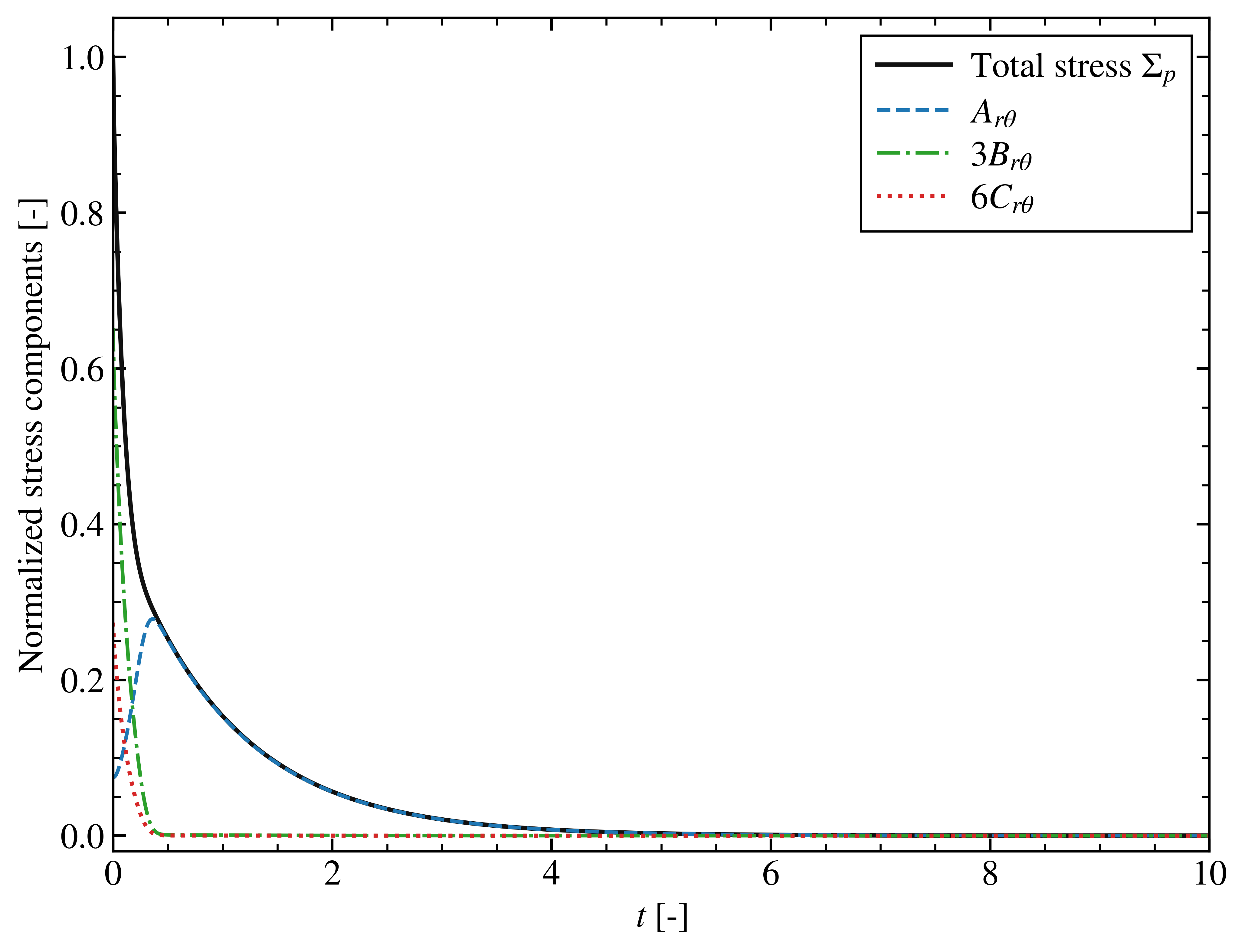}
    \caption{Microstructural stress partition during relaxation from a steady shear-banded state ($De=3.0$). The rapid stress release from the short and long species, coupled with the early-stage stress transfer to the newly formed gel-network, governs the multistep decay profile of the total macroscopic stress.}
    \label{fig:relaxation_stress_partition}
\end{figure}

\begin{figure*}[htbp]
    \centering
    \begin{subfigure}{0.32\textwidth}
        \centering
        \includegraphics[width=\linewidth]{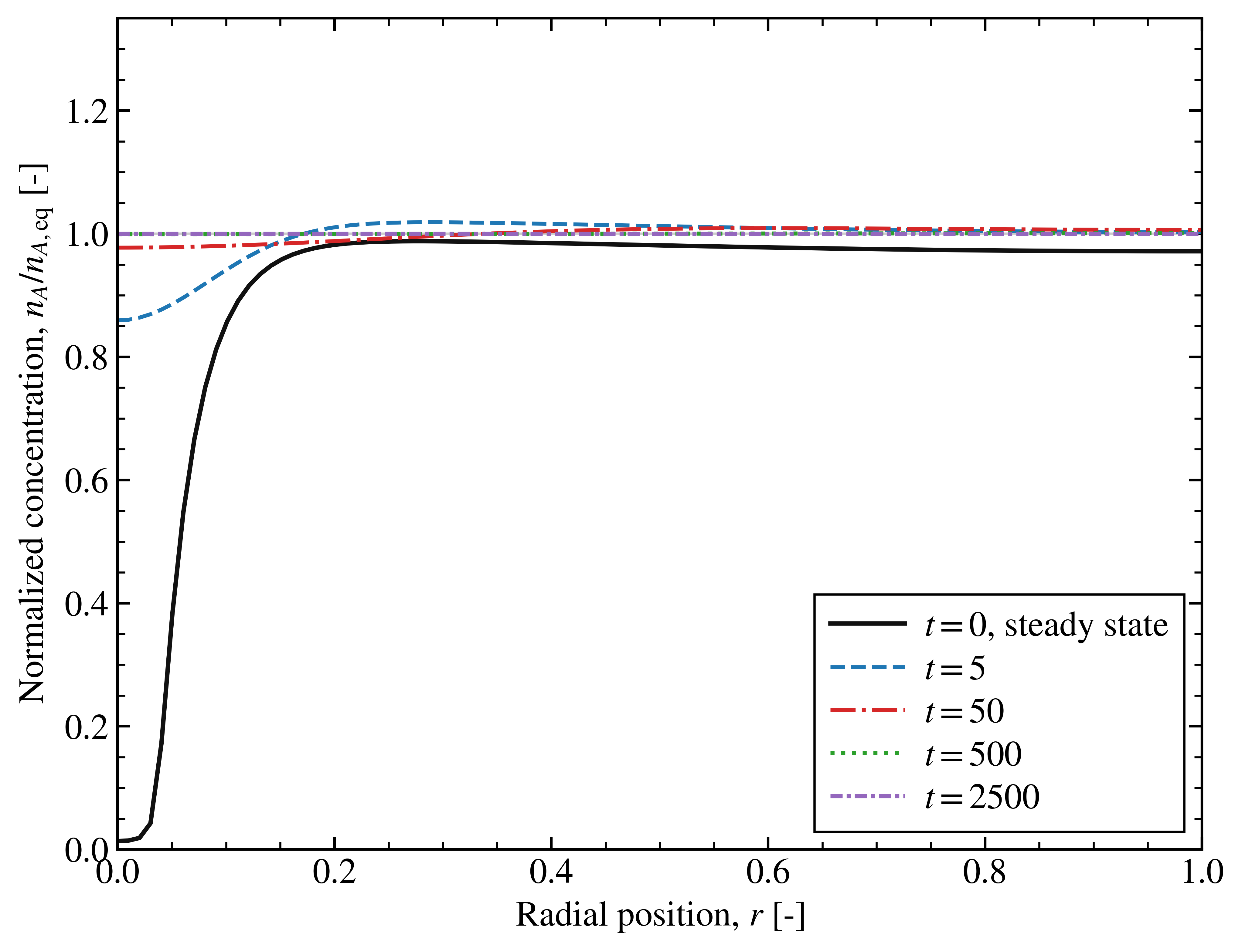}
        \caption{Spatial relaxation of $n_A$.}
    \end{subfigure}
    \hfill
    \begin{subfigure}{0.32\textwidth}
        \centering
        \includegraphics[width=\linewidth]{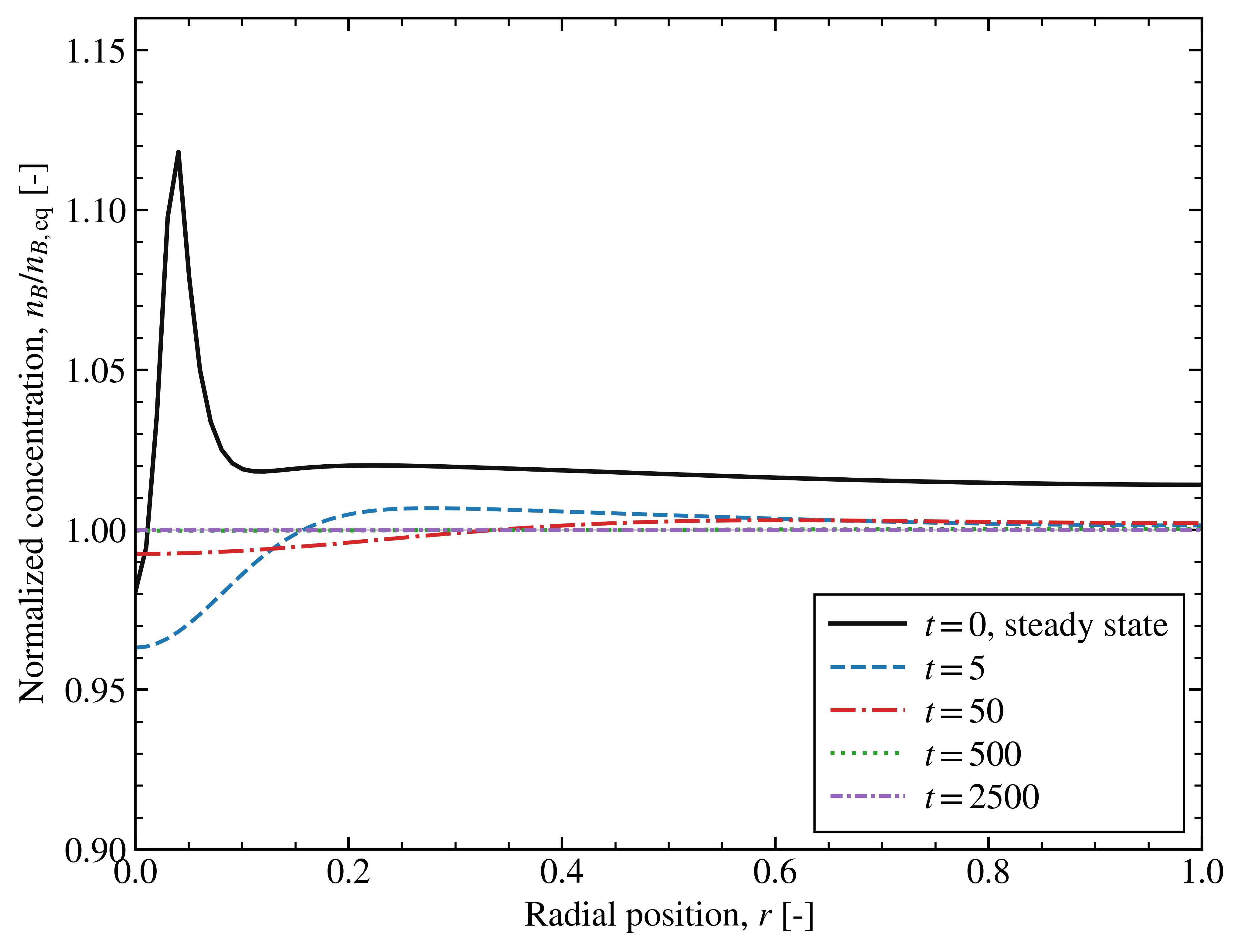}
        \caption{Spatial relaxation of $n_B$.}
    \end{subfigure}
    \hfill
    \begin{subfigure}{0.32\textwidth}
        \centering
        \includegraphics[width=\linewidth]{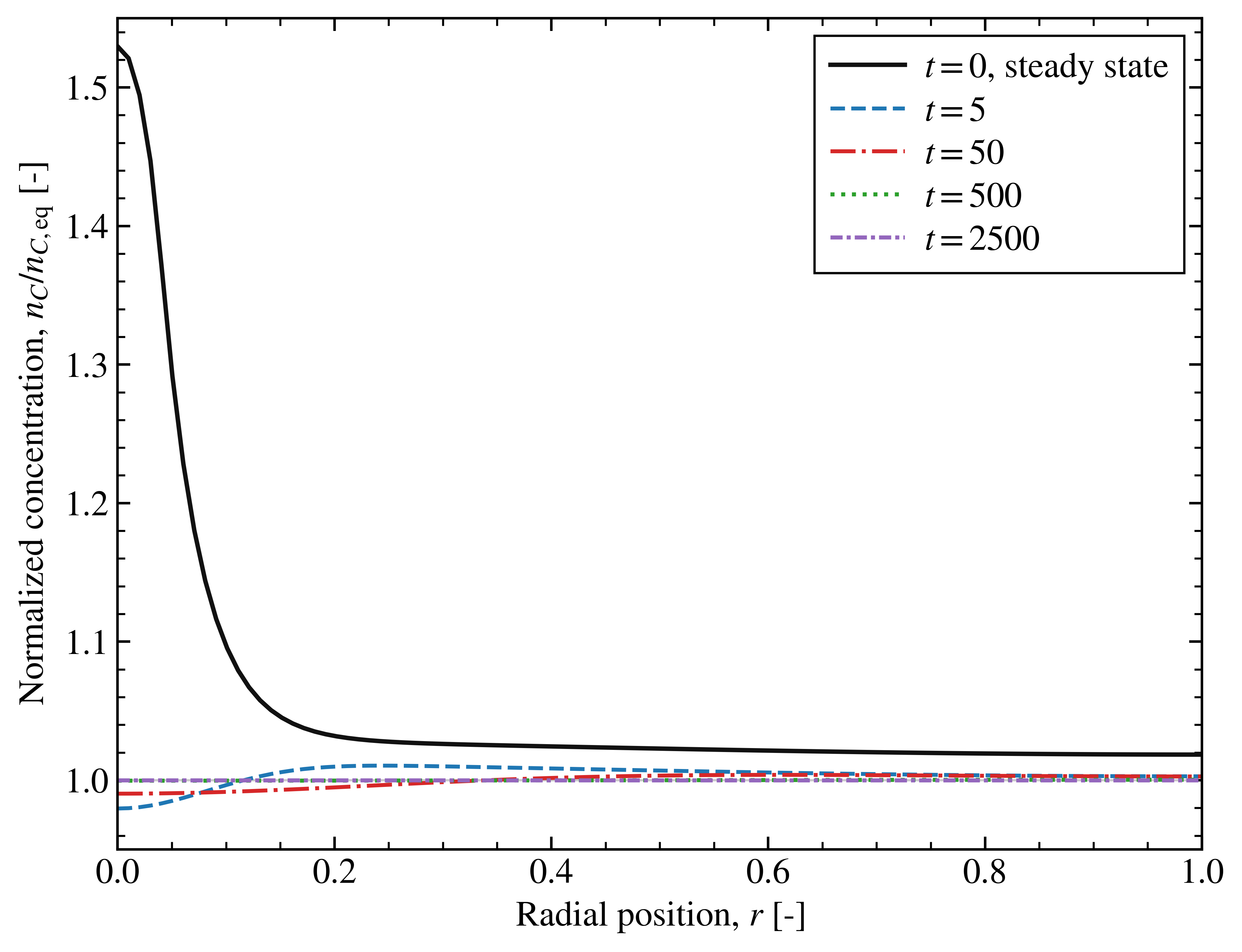}
        \caption{Spatial relaxation of $n_C$.}
    \end{subfigure}

    \caption{Spatial dynamics of the relaxing banded state ($De=3.0$) for the gel-network, long chains, and short chains, showing the gradual decay of the residual spatial imbalance from the highly inhomogeneous initial steady state at $t=0$ to the fully homogenized equilibrium state at $t=2500$.}
    \label{fig:relaxation_spatial_profiles}
\end{figure*}

The corresponding change in the stress carried by each structural level is presented in
Fig.~\ref{fig:relaxation_stress_partition}. Immediately after cessation, the total stress drops
rapidly because the elastic deformation stored in the shorter and more weakly sustained
structures is released first. At later times, the stress supported by the gel-network
becomes dominant in the remaining signal, and the overall decay slows down. Species B again plays an important role in connecting these two regimes. Its stress
contribution does not simply vanish at the earliest stage, nor does it persist as long as the
slowest network component. Instead, it occupies an intermediate time window and therefore
provides a direct structural explanation for the multistep form of the total stress relaxation.
This picture is also reflected in the spatial dynamics of the relaxing banded state. Representative
profiles in Fig.~\ref{fig:relaxation_spatial_profiles} show that, after cessation of shear, the sharp
contrast between the former high shear and low shear regions gradually weakens. The band
interface broadens, the velocity gradients disappear, and the concentration differences across
the gap slowly diminish. Nevertheless, this homogenization is not instantaneous. The memory
of the previously banded structure remains visible for a finite time, especially in species B and C, which retain a residual spatial imbalance even after the macroscopic
shear has been removed.

Overall, these results show that the present three-species cascade breakage
model naturally produces a multistep shear relaxation process. At the macroscopic level, this is
seen as a stress decay with clearly separated fast and slow stages. At the microstructural level,
it corresponds to a sequential redistribution among short chains, long chains, and gel-network species,
together with the gradual disappearance of the spatial heterogeneity generated during banding.
This behavior provides one of the clearest distinctions between the present model and simpler
reduced descriptions, since the additional intermediate structural level introduces an extra
relaxation channel and thereby broadens the overall relaxation spectrum.

\section{Conclusion}

In this work, we developed a three-species cascade breakage model to overcome the limitations of classical two-species scission frameworks in describing the multiscale rheology of wormlike micellar solutions. By explicitly resolving three distinct microstructural states, namely a primary gel-network, entangled long chains, and disentangled short chains, our model successfully recovers missing intermediate relaxation modes while maintaining macroscopic computational tractability.

Under homogeneous flow conditions, the proposed model analytically yields a three-mode Maxwell structure that accurately captures the high-frequency upturn in the dynamic moduli observed in experimental data. The framework also demonstrates rigorous continuum mechanical consistency by satisfying the Lodge-Meissner relation during large amplitude step strains. Driven by a flow-enhanced scission mechanism, the steady viscometric flow curve exhibits a pronounced non-monotonic response, where the physical parameters of the intermediate species strictly dictate the critical onset conditions for constitutive instability and shear banding.

When extended to inhomogeneous cylindrical Couette flows, the model captures the spontaneous formation of robust macroscopic shear bands, characterized by a clear spatial partition of the micellar populations. Crucially, species B naturally forms a microstructural transition layer across the macroscopic band interface. During transient start-up, the system exhibits classic stress overshoots driven by the competition between initial elastic loading and accelerated structural breakdown. Most notably, upon the cessation of shear, the presence of the intermediate state introduces an additional relaxation channel, resulting in a complex, multistep stress decay.

Ultimately, this three-species framework bridges the gap between discrete phenomenological network models and continuous kinetic theories. By resolving the cascade kinetics governing intermediate structural states, it provides a comprehensive physical basis for interpreting the rich transient and steady state dynamics of complex micellar fluids, offering new predictive capabilities for future microfluidic experiments and industrial applications.

\bibliographystyle{apsrev4-2}
\bibliography{paper}

\end{document}